\documentclass{iopjournal}

\usepackage[utf8]{inputenc}
\DeclareUnicodeCharacter{2009}{\,}
\usepackage{braket}
\usepackage{amsmath,amssymb}
\usepackage{bbm}
\usepackage{xcolor}
\usepackage{array}
\usepackage{colortbl}

\begin{document}

\articletype{Paper}

\title{Characterizing Phase Fragility via Algorithmically Prepared Ancillas in Repeated-Interaction Models}

\author{S. Elham Mousavigharalari$^1$ and Deniz T\"urkpen\c{c}e$^{1,2,*}$}

\affil{$^1$Informatics Institute, Istanbul Technical University, 34469 Maslak, \.{I}stanbul, T\"urkiye}

\affil{$^2$Qready Quantum Technologies, ITU ARI Teknokent, 34467 \.{I}stanbul, T\"urkiye}

\affil{$^*$Author to whom any correspondence should be addressed.}

\email{dturkpence@itu.edu.tr}


\begin{abstract}
We study how a phase parameter $\phi$, encoded through a single-qubit 
$H\,\varphi\,H$ gate sequence, is reflected in the quantum Fisher information 
(QFI) under realistic noisy dynamics. 
Within a collision-model framework, a probe qubit interacts sequentially with 
algorithmically prepared reservoir ancillas, leading to a $\phi$-dependent 
steady state from which $\mathcal{F}_{\phi}$ can be evaluated in closed form. 
In parallel, we perform pulse-resolved open-system simulations of the same 
gate sequence, using Gaussian-driven control motivated by transmon hardware, 
to obtain the corresponding pre-measurement density matrix. 
Despite the distinct physical descriptions, both approaches yield QFI 
profiles with similar phase dependence on the encoded phase. 
A quantitative comparison using profile-level similarity
metrics further shows that the two descriptions identify the same
phase-sensitive regions, although their absolute QFI contrasts differ.
This consistency indicates that the steady state of a probe qubit interacting 
with algorithmically prepared ancillas can capture key features of the phase 
response observed in noisy device-level implementations. 
The underlying physical mechanism is the persistence of finite 
steady coherence in the asymptotic probe state, which retains the phase imprint 
of the prepared ancillas.
Beyond conceptual insight, the steady-state framework provides a 
model-based, tomography-free diagnostic route 
for characterizing phase sensitivity. 
Possible uses in biased-noise error correction, 
hardware-aware compilation, and pulse-level optimization are therefore 
presented as future outlook directions.
\end{abstract}

\section{Introduction}
Today's quantum computers operate in the Noisy Intermediate Scale Quantum (NISQ) era~\cite{preskill_quantum_2018}. As a result, intensive efforts are underway, both in hardware and at the algorithmic level, to develop methods that allow quantum error correction (QEC) algorithms to function below their error threshold~\cite{shor_scheme_1995, knill_theory_1997, fowler_surface_2012, barends_superconducting_2014, ryan-anderson_realization_2021, acharya_quantum_2025}. Many quantum algorithms of practical importance, including Shor's factoring protocol~\cite{shor_algorithms_1994}, quantum phase estimation, and the Harrow–Hassidim–Lloyd solver~\cite{harrow_quantum_2009} encode their computational result in the relative phase $\phi$ between qubit states, typically in the form $\ket{\psi} \propto \ket{0} + e^{i\phi}\ket{1}$. 
Because this phase can only be read out through quantum interference, the encoded result is inherently vulnerable to dephasing processes. 
In current superconducting qubit architectures, dephasing often constitutes the dominant source of error, making the algorithmically relevant phase highly susceptible to noise during execution~\cite{zurek_decoherence_2003,mckay_three-qubit_2019}. This reality makes new approaches to quantifying the fragility of phase information under noisy hardware conditions highly relevant~\cite{len_quantum_2022, zhou_optimal_2023}. 

Quantum Fisher Information is a crucial mathematical tool in quantum metrology for estimating the precision of a quantum state's parameters~\cite{safranek_simple_2018, liu_quantum_2019}. However, calculating QFI requires knowledge of the state's density matrix ($\rho$), which in turn necessitates complex measurements like quantum state tomography (QST) across different bases. Even for a single basis measurement, a high number of experimental repetitions are needed to statistically average the outcomes. This process is complex, involving statistical effects such as shot noise inherent to measurement. These challenges motivate the development of approaches that assess phase-sensitive properties of quantum states directly from their noisy pre-measurement description.
 
The quantum collision model~\cite{scarani_thermalizing_2002, ziman_diluting_2002, nagaj_quantum_2002, cattaneo_collision_2021, karevski_quantum_2009, seah_nonequilibrium_2019, ciccarello_quantum_2022} has become a widely used framework for describing open quantum systems, primarily due to the flexibility it offers in parametrizing and engineering environmental degrees of freedom. In its conventional form~\cite{scarani_thermalizing_2002, ziman_diluting_2002, nagaj_quantum_2002, cattaneo_collision_2021}, a collision model emulates dissipative and Markovian open system dynamics through repeated, sequential interactions between identical reservoir units and a designated quantum probe. This repeated-interaction process, also referred to as quantum homogenization~\cite{ziman_diluting_2002, nagaj_quantum_2002}, typically drives the probe toward a stationary fixed point, which may correspond to thermal equilibrium or a non-equilibrium asymptotic state~\cite{karevski_quantum_2009, seah_nonequilibrium_2019}. In this regime, selected properties of the reservoir are dissipatively imprinted onto the probe system. Beyond standard thermal reservoirs, which are typically modeled by Gibbs-type mixed states, reservoir units can be prepared in parametrized mixed or pure states to represent structured, non-thermal environments~\cite{korkmaz_transfer_2022, ciccarello_quantum_2022,alves_collisional_2024}. This perspective is closely related to the concept of a quantum information reservoir~\cite{deffner_information_2013, deffner_information-driven_2013, korkmaz_quantum_2023}, where ancilla preparation plays an active role in shaping the effective open system dynamics.

This study is based on the working assumption that the pre--measurement state of a noisy quantum processor can, at an effective level, be viewed as the outcome of interactions with a structured and noisy environment. Within this perspective, we employ a collision model framework to investigate the interaction between a quantum probe qubit and an information reservoir composed of algorithmically prepared noisy ancilla units. In contrast to conventional thermal collision models, each ancilla in our setting is prepared through a finite--time noisy gate sequence, thereby acquiring a hardware specific noise imprint prior to interacting with the probe. The primary objective is to examine how parameters encoded in these ancilla states are transferred to, and reflected in, the asymptotic state reached by the probe under repeated interactions. Based on the analytical solution of the master equation describing this repeated--interaction dynamics~\cite{korkmaz_transfer_2022, korkmaz_quantum_2023}, we find that the probe state retains a parameter--dependent steady regime signature, which enables quantitative assessment of phase sensitivity. To prepare the reservoir ancillas, a gate sequence of $H\,\varphi\,H$ is employed, consisting of a Hadamard gate ($H$) and a phase shift gate ($\varphi$).

This gate sequence provides a device-independent, algorithmic representation
of the Mach--Zehnder interferometer~\cite{pezze_mach-zehnder_2008,
dowling_quantum_2008, giovannetti_advances_2011, pezze_quantum_2018}, a
paradigmatic tool in quantum metrology. In this context, the $H\,\varphi\,H$
circuit serves as a minimal algorithmic primitive for encoding relative phase
information under realistic noise.
This viewpoint is closely connected to earlier studies of
one-parameter qubit-gate estimation, including Bayesian and experimental
treatments of phase-gate estimation in noisy settings and in the presence of
phase diffusion~\cite{teklu_bayesian_2009, brivio_experimental_2010}. In the
present work, however, we do not estimate the gate parameter from measurement
data. Instead, we use the QFI of the pre-measurement state to characterize how
the noisy $H\,\varphi\,H$ primitive stores, transfers, and loses phase
sensitivity.
To this end, we pursue two complementary approaches:
(i) an exactly solvable repeated-interaction model in which a clean probe
qubit sequentially interacts with a stream of identically prepared noisy
ancillas, each of which has undergone the same $H\,\varphi\,H$ circuit under
decoherence, and
(ii) direct open-system simulation of the $H\,\varphi\,H$ sequence on a single
transmon qubit driven by Gaussian-shaped control pulses.

The novelty of this construction is that the reservoir units are not introduced as ideal thermal ancillas or as abstract parametrized mixed states. Instead, they are generated by a finite-time noisy quantum logic sequence and therefore carry an algorithmically prepared noise imprint before entering the
repeated-interaction dynamics. To the best of our knowledge, the present work
is the first to connect such an algorithmically prepared noisy ancilla reservoir
to a pulse-level open-system simulation of the same phase-encoding primitive,
thereby allowing the phase-sensitivity landscape of the reduced collision
description to be compared directly with that of a driven transmon-level
implementation.

Despite the fundamentally different physical mechanisms and mathematical descriptions underlying these two approaches, the resulting $\mathcal{F}_{\phi}$ profiles exhibit closely similar qualitative features, including the locations of maxima, minima, and zero crossings. This qualitative correspondence indicates that the repeated--interaction model captures key aspects of the phase sensitivity structure observed in the pulse--level device simulation. From a practical perspective, this suggests that the collision model analysis can serve as a lightweight, tomography--free diagnostic framework for exploring phase sensitivity trends in near term quantum hardware, without aiming to provide a fully microscopic description of the underlying device dynamics.

This manuscript is organized as follows. 
Section~\ref{Sect-pre} introduces the repeated interaction framework and formulates the probe--ancilla dynamics, together with a preliminary analysis based on mutual information. Section~\ref{Sect-model} presents the analytical evaluation of the probe’s quantum Fisher information using the asymptotic probe state obtained from the model. Section~\ref{Sect-analysis} describes pulse level, open system simulations of the $H\,\varphi\,H$ gate sequence on a transmon qubit and compares the resulting phase sensitivity trends with those of the repeated interaction analysis. Section~\ref{Sect-imp} discusses the implications of these results, and Section~\ref{Sect-conc} summarizes the main conclusions and outlines possible directions for future work.

\section{Preliminaries}\label{Sect-pre}

The quantum collision model, often referred to as a repeated--interaction framework, provides a versatile description of open quantum system dynamics in terms of sequential system--ancilla couplings. 
In this approach, a probe system interacts, one at a time, with a sequence of individually prepared ancillary units, each of which is traced out after a short interaction interval. 
The accumulation of these discrete interactions generates an effective dissipative evolution for the probe and can drive it toward an asymptotic fixed point over a finite number of interaction steps~\cite{turkpence_tailoring_2019, roman-ancheyta_spectral_2019}.

The dynamics of a single collision step can be precisely characterized by a quantum dynamical map, which is inherently completely positive and trace preserving (CPTP). This map, denoted here as $\Lambda$, transforms the system's state according to the following expression:

\begin{align}\label{Eq:map}
\Lambda[\rho_{S}] = \mathrm{Tr}_{\mathcal{R}}\!\Big[ U_{S\mathcal{R}} 
  \big(\rho_{S}\otimes\rho_{\mathcal{R}}\big) 
  U_{S\mathcal{R}}^{\dagger} \Big].
\end{align}

Here, $\rho_{S}$ represents the quantum state of the system of interest, and $\rho_{\mathcal{R}}$ is the initial state of a reservoir unit, which is typically prepared in a thermal or pure state. $U_{S\mathcal{R}}$ is the unitary operator describing the brief interaction between the system and the reservoir unit. The trace operation, $\text{Tr}_{\mathcal{R}}$, effectively averages over the degrees of freedom of the reservoir, yielding the updated state of the system.

When the reservoir is conceptualized as an information reservoir, it is often convenient to examine the quantum mutual information between the probe system and the reservoir units. The mutual information that quantifies how much information has flowed from the reservoir unit to the probe is
\begin{equation}\label{Eq:Mutual}
\mathcal{I}\left( S:R_i\right)=S\left(\rho_{S}^{(i)}\right)+S\left(\rho_{R_i}\right)
-S\left(\rho_{SR_i}^{(i)}\right)
\end{equation}
where  
\(S(\rho)=-\operatorname{Tr}[\rho\log_2\rho]\) is the von Neumann entropy.
Here, $\rho_{SR_i}^{(i)}$ is the joint density matrix immediately after the $i$-th collision, $\rho_{S}^{(i)}=\operatorname{Tr}_{R_i}\rho_{SR_i}^{(i)}$ is the updated probe state and $\rho_{R_i}=\operatorname{Tr}_{S}\rho_{SR_i}^{(i)}$ is the outgoing ancilla state that is subsequently traced out.

The precision of parameter estimation is commonly characterized by the Fisher information. 
For a classical distribution $\{p_r(\lambda)\}$, it is defined as
\begin{equation}
\mathcal{F}_{\lambda}=\sum_r p_r(\lambda)\left[\frac{\partial }{\partial\lambda}\ln p_r(\lambda)\right]^2,
\end{equation}
which leads directly to the Cramér--Rao inequality 
$\mathrm{Var}(\hat{\lambda}) \geq [M\,\mathcal{F}_{\lambda}]^{-1}$ for any unbiased estimator $\hat{\lambda}$ based on $M$ repetitions.

In the quantum setting, one considers a family of density operators $\rho_{\lambda}$ and defines the quantum Fisher information (QFI) through the symmetric logarithmic derivative $L_{\lambda}$, implicitly determined by
$\partial_{\lambda}\rho_{\lambda} = \tfrac{1}{2}(\rho_{\lambda} L_{\lambda} + L_{\lambda}\rho_{\lambda})$~\cite{helstrom_quantum_1969}. 
The resulting quantum Cramér--Rao bound,
\begin{equation}
\mathrm{Var}(\hat{\lambda}) \geq \frac{1}{M\,\mathcal{F}_{\lambda}},
\end{equation}
sets the ultimate sensitivity allowed by quantum mechanics~\cite{braunstein_statistical_1994}.
For two-level systems, several equivalent forms of the QFI are known~\cite{dittmann_explicit_1999,zhong_fisher_2013}.  
Since our analysis involves a single qubit, it is convenient to work with the Bloch representation 
$\rho = \tfrac{1}{2}(\mathbb{I} + \mathbf{r}\!\cdot\!\boldsymbol{\sigma})$, 
where $\mathbf{r}$ is the Bloch vector.  
In this form, the QFI takes the compact expression
\begin{align}\label{Eq:Fisher_Bloch}
\mathcal{F}_{\lambda} 
    = |\partial_{\lambda}\mathbf{r}|^{2}
    + \frac{\bigl(\mathbf{r}\cdot\partial_{\lambda}\mathbf{r}\bigr)^{2}}{1 - |\mathbf{r}|^{2}},
\end{align}
which is well suited for mixed states with $|\mathbf{r}|<1$ and reduces to 
$\mathcal{F}_{\lambda}=|\partial_{\lambda}\mathbf{r}|^{2}$ in the pure-state limit.

The phase sensitivity analyzed in this work is extracted from the density matrix of the probe qubit obtained after repeated interactions with an information reservoir. In the collision model description, this corresponds to the asymptotic probe state reached after many sequential interactions with identically prepared ancilla units. Our primary objective is to characterize the quantum Fisher information of this probe state when the reservoir units are algorithmically prepared and carry a finite time noise signature. To place these results in a hardware-relevant context, we compare the collision model prediction with the quantum Fisher information obtained from a direct device-level simulation of the same $H\,\varphi\,H$ gate sequence executed on a noisy single qubit platform. Specifically, the QFI is evaluated from two complementary descriptions: (i) the asymptotic probe state generated by the collision model and (ii) the output density matrix obtained from a pulse-level open system simulation of the $H\,\varphi\,H$ circuit.

The device-level dynamics are modelled within a standard open system framework~\cite{breuer_theory_2010}, where the driven evolution of the qubit is described by a time-dependent Lindblad master equation. The density operator obeys
\begin{align}\label{Eq:Lindblad}
\frac{d}{dt}\rho(t) = -\frac{i}{\hbar}[H(t),\rho(t)] 
+ \sum_k \gamma_k \Big(L_k \rho(t) L_k^\dagger - \tfrac{1}{2}\{L_k^\dagger L_k,\rho(t)\}\Big),
\end{align}
which accounts for both coherent control and dissipative processes. For time-dependent control Hamiltonians, the evolution is obtained by numerically integrating this equation, which is equivalent to applying a time-ordered exponential of the associated Liouvillian superoperator. This formulation is standard in pulse-level simulations of driven superconducting qubits and captures the effects of relaxation and dephasing during gate execution~\cite{motzoi_simple_2009, chow_optimized_2010}.

\begin{figure}[t!]
\centering
\includegraphics[width=5.2in]{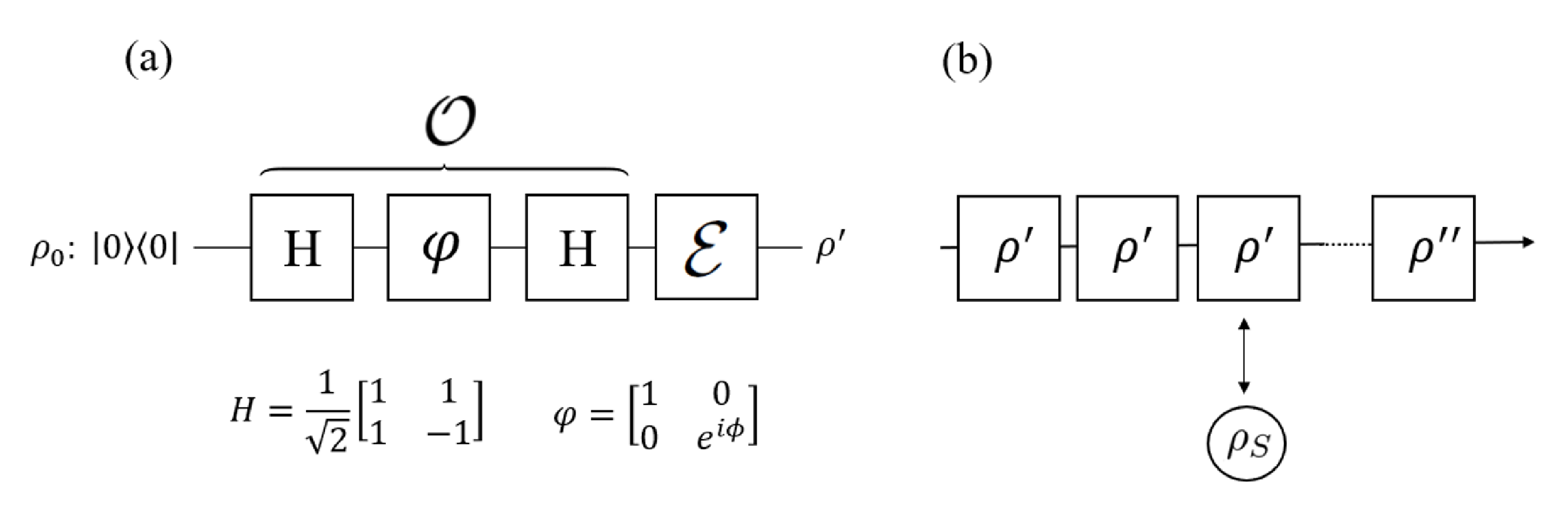}
\caption{ \label{fig:Fig1} Schematic of the proposed protocol.  
(a) Algorithmic and noisy preparation of reservoir qubits: each unit is 
initialized to $\rho_0$, evolved under the unitary sequence $H\,\varphi\,H$, 
and subsequently propagated for a finite preparation time through a 
phenomenological noise channel $\mathcal{E}$, yielding the mixed state 
$\rho'$.  (b) Sequential interaction dynamics within a collision model framework: 
the sequence of identically prepared ancillas, each carrying a finite time 
noise signature, collectively acts as a single effective information 
reservoir.  The ancillas are injected one by one, each interacting briefly with the probe qubit before being discarded, thereby imprinting $\phi$-dependent information 
onto the asymptotic state of the probe~\cite{scarani_thermalizing_2002,ciccarello_quantum_2022}.
} 
\end{figure}

\section{Model and system dynamics}\label{Sect-model}

The fragility of valuable quantum information under realistic noisy conditions
has motivated researchers to better understand the role of decoherence and
dissipation not only as sources of information loss, but also as mechanisms
that can mediate structured information transfer~\cite{verstraete_quantum_2009}.
In this broader perspective, quantum reservoirs need not merely function as
irreversible sinks for system information, but can instead act as structured
communication channels, transferring specific signatures of the environment to
the system~\cite{seah_nonequilibrium_2019, blume-kohout_simple_2005,
zwolak_redundancy_2017}. 
This viewpoint is also conceptually related to studies of
noisy quantum phase communication channels, where information is encoded in a
phase shift and degraded by phase diffusion or related noise mechanisms
~\cite{teklu_noisy_2015, trapani_quantum_2015, adnane_quantum_2019}. Although
those works address communication-channel settings rather than the
collision-model framework considered here, they provide a relevant precedent
for connecting noisy phase encoding, phase fragility, and
information-based characterization.

In this context, the quantum collision model provides a convenient and analytically tractable framework for tracking how parameter-dependent features of an engineered information reservoir are transferred to a probe system through repeated interactions. In particular, the model allows one to characterize the asymptotic probe state reached after many sequential interactions, which we refer to as a steady state probe response in the sense of a fixed point of the repeated interaction dynamics. This tractability enables the derivation of effective master equations that allow direct evaluation of the quantum Fisher information associated with this asymptotic probe state.

Fig.~\ref{fig:Fig1} provides an overview of the proposed framework, illustrating 
(a) the noisy, finite time preparation of individual reservoir units and 
(b) their sequential interactions with a probe system within a collision model. 
In this scheme, both the reservoir units and the probe system are modeled as 
two-level quantum systems (qubits).

The information reservoir qubits are prepared by first applying the single–qubit 
gate sequence \(\mathcal{O}=H\varphi H\), with \(H\) the Hadamard gate and 
\(\varphi\) a relative-phase rotation, and subsequently evolving the output 
state for a finite preparation time under a phenomenological noise channel 
\(\mathcal{E}\). 
The net noisy preparation map is therefore 
\(\widetilde{\mathcal{O}}=\mathcal{E}\!\circ\!\mathcal{O}\), which acts on the 
initial state \(\rho_0\) to produce the identical, non-interacting reservoir 
state
\[
\rho' \;=\;\widetilde{\mathcal{O}}\,[\rho_0]
       \;=\;\mathcal{E}\!\bigl(\mathcal{O}\rho_0\mathcal{O}^{\dagger}\bigr).
\]

In this study, the information reservoir is explicitly defined as a single 
effective environment simulated by a collection of $n$ uncorrelated, identical 
qubit states (ancillas), each of which carries an algorithmically prepared, 
finite time noise signature prior to its interaction with the probe. 
Collectively, the reservoir is represented by the tensor product
\begin{align}\label{Eq:InfRes}
\rho_{\mathcal{R}}=\bigotimes_{i=1}^n\rho'_{i}(\phi).
\end{align}
This construction follows the standard collision model framework, where a 
sequence of ancillas is used to discretize and emulate the continuous interaction 
with a single, Markovian or non-Markovian, external bath~\cite{scarani_thermalizing_2002, ciccarello_quantum_2022}, 
while extending it to the case of algorithmically prepared, non-thermal 
reservoir units.

Before writing the explicit noisy ancilla state, we specify the noise
convention used in the finite preparation stage. Starting from the initial
state $\ket{0}$, the ideal $H\,\varphi\,H$ sequence produces the density matrix
\begin{equation}
\rho\sb{\mathrm{id}}(\phi)
=
\begin{pmatrix}
\frac{1+\cos\phi}{2} & \frac{i\sin\phi}{2} \\[4pt]
-\frac{i\sin\phi}{2} & \frac{1-\cos\phi}{2}
\end{pmatrix}.
\label{Eq:Rho_ideal_ancilla}
\end{equation}
During the finite preparation interval $t\sb{p}$, this state is subjected to a
phenomenological single-qubit noise map consisting of longitudinal population
relaxation and transverse coherence damping. In the ordered basis
$\{\ket{0},\ket{1}\}$ used here, the relaxation map is written such that the
population associated with the first basis component relaxes into the second
basis component. Thus,
\begin{align}
\rho\sb{00}(t\sb{p}) &= \rho\sb{00}(0)\exp(-\gamma\sb{1}t\sb{p}), \\
\rho\sb{11}(t\sb{p}) &= 1-\rho\sb{00}(0)\exp(-\gamma\sb{1}t\sb{p}), \\
\rho\sb{01}(t\sb{p}) &= \rho\sb{01}(0)\exp(-\gamma\sb{2}t\sb{p}), \\
\rho\sb{10}(t\sb{p}) &= \rho\sb{10}(0)\exp(-\gamma\sb{2}t\sb{p}).
\end{align}
Here, $\gamma\sb{1}=1/T\sb{1}$ is the amplitude-relaxation rate and
$\gamma\sb{2}=1/T\sb{2}$ is the effective transverse decoherence rate. The
latter may include both relaxation-induced coherence loss and pure dephasing
contributions. This convention yields a trace-preserving finite-time prepared
ancilla state and fixes the basis ordering used in the following expression.
Applying this map to Eq.~\eqref{Eq:Rho_ideal_ancilla} gives the prepared noisy
reservoir unit used throughout the repeated-interaction analysis:
The quantum state of each noisy reservoir unit, which is prepared by applying phenomenological noise to the algorithmic protocol in Fig.~\ref{fig:Fig1}(a), can be straightforwardly shown to be:

\begin{equation}\label{Eq:Rho_noise}
\rho'(\phi,t_p) = 
\begin{bmatrix}
\frac{1+\cos\phi}{2}e^{-t_p/T_1} 
& \frac{i \sin\phi}{2}e^{-t_p/T_2} \\[4pt]
\frac{-i \sin\phi}{2}e^{-t_p/T_2} 
& 1 - \frac{1+\cos\phi}{2}e^{-t_p/T_1}
\end{bmatrix}.
\end{equation}

Here, $\rho'(\phi,t_p)$ denotes the algorithmically and dissipatively prepared 
state of each reservoir ancilla prior to its interaction with the probe. 
The time $t_p$ represents a finite noise–preparation interval during which 
each ancilla is exposed to relaxation and dephasing channels characterized 
by $T_1$ and $T_2$. Importantly, $t_p$ is not taken to the asymptotic limit, but instead corresponds to a finite hardware-relevant 
preparation time. In this sense, the reservoir units are \emph{frozen} in a mixed state that encodes a finite time noise signature before entering the repeated–interaction dynamics. The subsequent collision dynamics therefore does not model 
continuous dissipation acting on the reservoir units during the interaction, 
but rather the transfer of information from algorithmically prepared, 
noise-imprinted ancillas to the probe system. This construction defines an 
engineered information reservoir in which finite time noisy preparation, 
rather than infinite-time thermalization, determines the effective reservoir 
state.

The overall quantum dynamical map for a sequence of $n$ collisions, taking place 
over a total time of $n\tau$, can be expressed as a nested composition of trace 
operations, which describes the sequential interaction of the probe system 
with each reservoir unit prepared in the finite time noisy state 
$\rho'(\phi,t_p)$:
\begin{align}
\Lambda_{n\tau}[\rho_S]=\text{Tr}_n \big[ U_n \ldots 
\text{Tr}_1[U_1\left(\rho_S\otimes\rho'_{1}\right)U_1^{\dagger}]
\otimes\ldots 
\ldots\otimes\rho'_{n} U_n^{\dagger} \big].
\end{align}

Here, $\tau$ represents the duration of a single collision, and 
$U_i=e^{-i\mathcal{H}_i\tau}$ is the unitary propagator for the $i$-th interaction 
step. The total Hamiltonian for this interaction is given by 
$\mathcal{H}_i=\mathcal{H}_{S,i}^{\text{free}}+\mathcal{H}_{S,i}^{\text{int}}$, 
where the free term and the interaction term are defined as:
\begin{align}
\mathcal{H}_{S,i}^{\text{free}} &= 
\frac{\hbar\omega_S}{2}\sigma_S^z
+\frac{\hbar\omega_i}{2}\sigma_i^z, \\
\mathcal{H}_{S,i}^{\text{int}} &= 
\hbar g(\sigma_S^{+}\sigma_i^{-}+\text{h.c.}).\label{Eq:HAM}
\end{align}

In these expressions, $\sigma_S^z$ and $\sigma_i^z$ are the Pauli-$z$ operators 
for the probe system and the $i$-th reservoir unit, respectively, with 
corresponding frequencies $\omega_S$ and $\omega_i$. In general, we take 
$\omega_S=\omega_i$ for each collision step. The operators $\sigma^{+}$ and 
$\sigma^{-}$ denote the standard raising and lowering operators, and $g$ is the 
coupling constant characterizing the strength of the partial-swap interaction.

During each collision step, the probe interacts coherently with a 
\emph{frozen, algorithmically prepared} ancilla whose mixed state encodes 
a finite time noise signature acquired prior to the interaction. 
No additional dissipative dynamics is assumed to act on the reservoir units 
during the collision itself; instead, all dissipation and dephasing affecting 
the reservoir are incorporated through the finite preparation time $t_p$ in 
$\rho'(\phi,t_p)$. The non-unitary evolution of the probe therefore arises 
solely from the repeated interaction and partial trace over successive ancillas.

\subsubsection{Collisional Mutual Information}

To characterize how parametric information is transferred from an engineered 
information reservoir to the probe, we first analyze the mutual information 
generated during the repeated interaction process. In particular, the 
$\phi$-dependent evolution of the mutual information between the probe qubit 
and successive reservoir ancillas provides a quantitative indicator of how 
algorithmically encoded phase information is imprinted through the collision 
dynamics.

\begin{figure}[!t]
\centering
\includegraphics[width=4.0in]{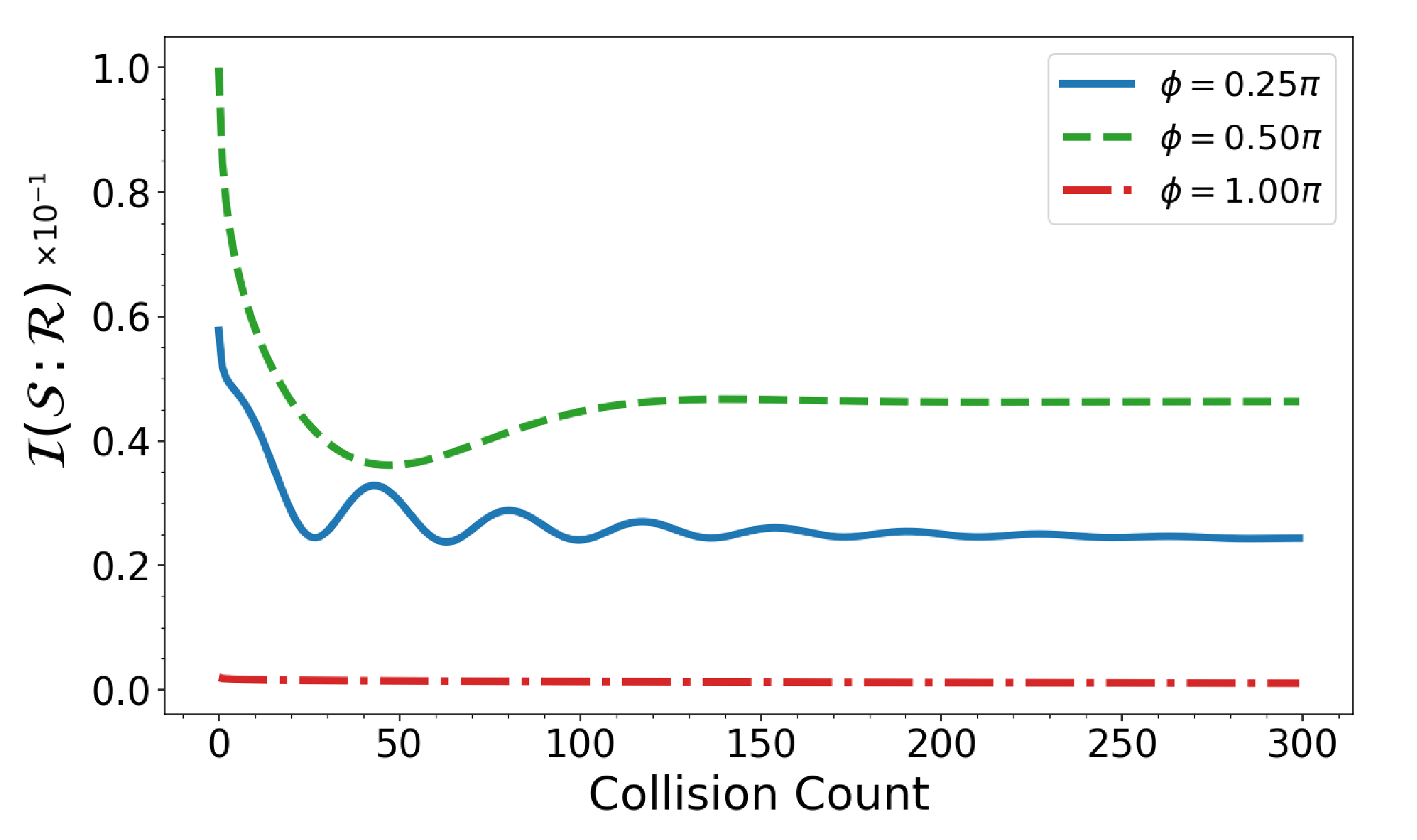}
\caption{ \label{fig:Fig2} Mutual information between the probe qubit and the sequence of reservoir ancillas as a function of the collision count for different values of the phase parameter $\phi$. Each ancilla is prepared by a finite time noisy preparation stage characterized by phenomenological relaxation and dephasing 
times $T_1=150~\mu\mathrm{s}$ and $T_2=100~\mu\mathrm{s}$, which determine the 
coherence and population imbalance encoded prior to the collision. The probe 
and reservoir level splittings are set to 
$\omega_{\mathcal S}=\omega_{\mathcal R}\equiv 1$. The probe--ancilla coupling 
strength is $g=0.1$ and the collision duration is $\tau=0.12$, both given in 
units of $1/\omega_{\mathcal S}$.} 
\end{figure}

Fig.~\ref{fig:Fig2} displays the mutual information between the probe qubit 
and the phase-parameterized information reservoir as a function of the collision 
count. Each reservoir ancilla is prepared by applying the algorithmic protocol 
shown in Fig.~\ref{fig:Fig1}(a), followed by a finite time exposure to 
phenomenological dissipation and dephasing channels. This preparation results 
in a mixed ancilla state $\rho'(\phi,t_p)$ that carries a controlled, 
finite time noise signature prior to the interaction with the probe.

Numerical simulations were performed with the \textsc{QuTiP} library~\cite{johansson_qutip_2013}, 
adopting the convention $\hbar=1$ throughout. For the dissipation and dephasing 
times entering the noisy preparation map, we chose characteristic 
superconducting–qubit relaxation parameters $T_1$ and $T_2$ on the order of 
$\mu\mathrm{s}$. The noise–preparation interval of each reservoir ancilla was 
set to $t_p=480\,\text{ns}$, corresponding to the execution time of several 
single–qubit logic gates in current hardware~\cite{ibm_q_experience_2025}. 
Importantly, this finite preparation time defines the degree of decoherence 
encoded in each ancilla prior to the collision, while no additional 
dissipative evolution of the ancillas is assumed during the interaction itself.

The probe is initialized in the superposition state 
$\ket{+}=(\ket{0}+\ket{1})/\sqrt{2}$ in order to maximize its sensitivity to 
phase-dependent coherence transferred from the reservoir. As shown in 
Fig.~\ref{fig:Fig2}, the mutual information $\mathcal{I}(S{:}\mathcal R)$ 
generated between the probe and the reservoir exhibits a pronounced dependence 
on the reservoir phase $\phi$ and relaxes toward distinct asymptotic values 
for different phase settings. The smallest asymptotic value occurs for 
$\phi=\pi$, whereas the largest is observed for $\phi=\tfrac{\pi}{2}$.

This behavior can be understood from the structure of the finite time prepared 
ancilla states. The transverse coherence encoded in $\rho'(\phi,t_p)$ scales 
with $e^{-t_p/T_2}|\sin\phi|$, while the diagonal populations are modified by 
$T_1$ relaxation in a trace-preserving manner. Since the exchange-type 
interaction generates correlations predominantly through the coherent 
$\{\ket{01},\ket{10}\}$ manifold, the mutual information accumulated in the 
probe reflects this coherence envelope. Consequently, correlation build-up is 
enhanced near $\phi=\tfrac{\pi}{2}$, where the ancilla coherence is maximal and 
populations are approximately balanced, and suppressed near $\phi=\pi$ (and 
$\phi\approx 0$), where the ancilla state becomes effectively 
$\sigma_z$–eigenlike and therefore weakly coupled for information transfer.

\subsubsection{Micromaser Master Equation}

Micromaser systems~\cite{filipowicz_theory_1986,cresser_quantum-field_1992}
have long provided a physically transparent route for deriving effective
continuous-time master equations from sequences of discrete
system--ancilla interactions~\cite{liao_single-particle_2010,turkpence_quantum_2016}.
In the present context, this formalism is adopted as a coarse-grained
description of a repeated-interaction process in which a single probe qubit
sequentially couples to a stream of identically prepared ancillas, each in
the finite time prepared noisy state $\rho'(\phi)$, before being discarded.

In the interaction picture, the propagator for a single collision admits the
second-order expansion in the short interaction time $\tau$,
\begin{align}
U(\tau) \simeq 1 - i\tau V - \tfrac{\tau^{2}}{2} V^{2},
\qquad 
V = g\bigl(\sigma^{+}_{S}\sigma^{-}_{i} + \sigma^{-}_{S}\sigma^{+}_{i}\bigr),
\end{align}
which is standard in micromaser and repeated-interaction treatments.

To derive a time-continuous master equation, we follow the conventional
micromaser coarse-graining procedure, in which collisions are assumed to occur
according to a Poisson process with average rate $r$ (see Appendix~\ref{App:A}
for details). This assumption provides a convenient mathematical device for
passing from the discrete collision picture to a Lindblad-form generator.
Importantly, in all numerical and analytical evaluations presented below, we
fix $r=1$, which corresponds to a rescaling of the time variable and does not
affect the structure of the steady or asymptotic dynamics. The resulting
equation therefore also applies, up to a trivial time rescaling, to uniform
sequential interactions.

The resulting effective master equation for the reduced probe state reads
\begin{align}\label{eq:micro_me_final}
\dot{\rho}_S(t)
= -i\,[H_{\mathrm{eff}},\rho_S(t)]
   + \Gamma_{+}\,\mathcal{L}[\sigma^{+}_{S}](\rho_S(t)) + \Gamma_{-}\,\mathcal{L}[\sigma^{-}_{S}](\rho_S(t)),
\end{align}
with effective Hamiltonian and rates
\begin{align}
H_{\mathrm{eff}} &= r\,\tau\,g\!\left(\langle\sigma^{-}\rangle_{\rho'}\,\sigma^{+}_{S}
                  + \langle\sigma^{+}\rangle_{\rho'}\,\sigma^{-}_{S}\right), \\[2mm]
\Gamma_{+} &= \tfrac{1}{2}r\,\tau^{2}g^{2}\,\langle\sigma^{+}\sigma^{-}\rangle_{\rho'}, \qquad
\Gamma_{-} = \tfrac{1}{2}r\,\tau^{2}g^{2}\,\langle\sigma^{-}\sigma^{+}\rangle_{\rho'}.
\end{align}

All coefficients are expressed as expectation values over the ancilla state
$\rho'(\phi)$, making explicit how the algorithmically prepared, finite time
noisy ancillas shape both the coherent and dissipative contributions to the
probe dynamics.

Setting $\dot{\rho}_S=0$ yields the corresponding asymptotic fixed point of the
coarse-grained dynamics,
\begin{align}\label{Eq:Steady_rho}
\rho_{S}^{\mathrm{ss}} 
= \langle\sigma^{+}\sigma^{-}\rangle_{\rho'}\,|0\rangle\langle 0|
 +  \langle\sigma^{-}\sigma^{+}\rangle_{\rho'}\,|1\rangle\langle 1| 
 + \bigl[
i\,\gamma^{-}\!\left(
  \langle\sigma^{+}\sigma^{-}\rangle_{\rho'}
 -\langle\sigma^{-}\sigma^{+}\rangle_{\rho'}
\right)|0\rangle\langle 1| + \text{H.c.} \bigr],
\end{align}
where $\gamma^{-}=r\,\tau\,g\,\langle\sigma^{-}\rangle_{\rho'}$.

\section{Analytical and numerical analysis}\label{Sect-analysis}

In this section, we present a comparative study between the analytical expressions derived from the steady state solution of the micromaser master equation and the numerical evaluation of quantum Fisher information (QFI) obtained from device level noisy quantum simulations. The analytical results highlight how phase sensitivity is encoded in the asymptotic density matrix, while the numerical analysis employs circuit–level noise models   to capture hardware–specific imperfections. By aligning these approaches, we aim to establish a consistent framework that links theoretical predictions with realistic experimental scenarios.

\subsubsection{Quantum Fisher Information}

Within the repeated--interaction framework, phase information encoded in the
prepared reservoir units is transferred to the probe through successive
dissipative probe--ancilla interactions. The resulting asymptotic probe state,
denoted $\rho_{S}^{\mathrm{as}}(\phi)$, follows from the effective micromaser
master equation and is given in Appendix~\ref{App:B}. 
Although this asymptotic state remains mixed due to relaxation and dephasing,
finite off-diagonal elements persist. This non-vanishing steady coherence is
the physical property that allows the probe to retain information about the
encoded phase in the long-time regime~\cite{karevski_quantum_2009, seah_nonequilibrium_2019, korkmaz_quantum_2023, huang_steady-state_2025}. In particular, the transverse coherence
initially present in the algorithmically prepared ancillas is transferred to
the probe through repeated interactions and appears as a finite coherence term
in $\rho\sb{S}\sp{\mathrm{ss}}(\phi)$. The metrological response of the probe,
as quantified by $\mathcal{F}\sb{\phi}$, therefore follows directly from the
properties of the asymptotic probe state rather than from the comparison with
the device-level simulation alone.

\begin{figure}[!t]
\centering
\includegraphics[width=4.0in]{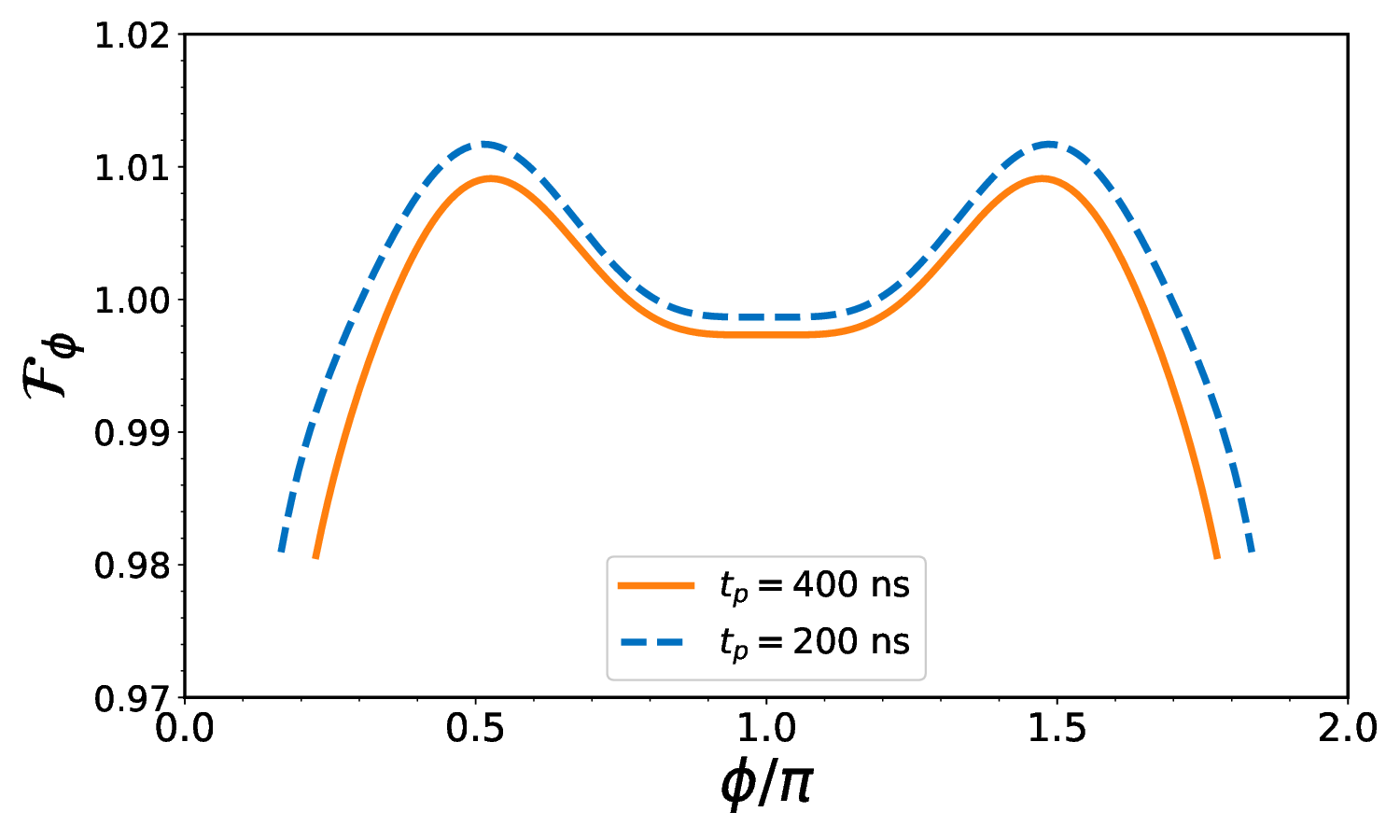}
\caption{ \label{fig:Fig3} Quantum Fisher information $\mathcal{F}_{\phi}$ of the probe qubit as a function of the encoded phase $\phi$ for two finite noise--preparation intervals, $t_p=200~\mathrm{ns}$ and $t_p=400~\mathrm{ns}$. Dissipative effects are incorporated through the phenomenological relaxation constants $T_{1}=150~\mu\mathrm{s}$ and $T_{2}=100~\mu\mathrm{s}$. The probe--reservoir interaction scale is fixed by the dimensionless parameter $\zeta=0.12$, which corresponds to an interaction strength $g=0.15$ and a collision duration $\tau=0.8$ in units of $1/\omega_{\mathcal S}$.} 
\end{figure}

Instead of working directly with the density matrix, it is convenient to use
the Bloch representation
$\boldsymbol{r}(\phi)=\bigl(r_x(\phi),\,r_y(\phi),\,r_z(\phi)\bigr)$,
obtained from the standard relations
$r_i=\mathrm{Tr}\!\left[\rho_{S}^{\mathrm{as}}\sigma_i\right]$.
The explicit expressions for the components $r_x(\phi)$ and $r_z(\phi)$,
together with their derivatives, are derived in Appendix~\ref{App:C}.
These expressions incorporate the dissipative parameters
$\gamma_1=1/T_1$ and $\gamma_2=1/T_2$, as well as the effective coupling
strength $\zeta = r\,\tau\,g$ that characterizes the probe--reservoir
interaction.
Substituting these results into the Bloch--vector expression for the quantum
Fisher information yields the closed analytical form in
Eq.~\eqref{Eq:QFI-closed}, which is used throughout the following analysis.

The dependence of the QFI on $\phi$ reflects the competition between the
longitudinal component $r_z(\phi)$, which is suppressed by energy relaxation,
and the transverse component $r_x(\phi)$, which carries the phase dependence
but is attenuated by both relaxation and dephasing.
As shown in Fig.~3, this interplay produces a characteristic oscillatory
structure with maxima around $\phi=\pi/2$ and $3\pi/2$, and a minimum near
$\phi=\pi$.

The two curves in Fig.~3 correspond to different finite noise--preparation
intervals, $t_p=200~\mathrm{ns}$ and $t_p=400~\mathrm{ns}$, which set the
effective noise imprint carried by each reservoir ancilla prior to its
interaction with the probe.
For the longer preparation interval ($t_p=400~\mathrm{ns}$), the QFI amplitude
is slightly reduced, reflecting the increased attenuation of transverse
coherence during the ancilla preparation stage.
This behavior highlights that the asymptotic phase sensitivity depends not only
on the interaction dynamics, but also on the finite time noise processes that
shape the prepared reservoir states.

A comparison with the mutual information behavior shown in
Fig.~\ref{fig:Fig2} reveals a closely related structure.
In both cases, the dependence on $\phi$ is governed by the same
reservoir--induced coherence profile: maximal values occur near
$\phi=\pi/2$, while a minimum appears around $\phi=\pi$.
Although mutual information quantifies total probe--reservoir correlations and
QFI captures local phase sensitivity, the locations of their extrema coincide.
This correspondence indicates that operating points associated with stronger
steady probe--reservoir correlations also coincide with enhanced sensitivity of
the probe to the encoded phase parameter. 
We use different interaction parameters in Figs.~\ref{fig:Fig2} and
\ref{fig:Fig3} because the two figures serve different diagnostic purposes.
In Fig.~\ref{fig:Fig2}, the aim is to resolve the transient build-up of mutual
information over many collisions. We therefore choose a weak effective
interaction, $g=0.10$ and $\tau=0.12$, corresponding to $g\tau=0.012$, so that
the correlation dynamics does not saturate within only a few collision steps.
In Fig.~\ref{fig:Fig3}, the focus is instead on the steady-state QFI profile
as a function of the encoded phase. For this purpose we use
$\zeta=g\tau=0.12$, with $g=0.15$ and $\tau=0.80$, which makes the
phase-dependent modulation of $\mathcal{F}_{\phi}$ more clearly visible while
remaining within the perturbative regime of the micromaser approximation. Thus,
the different parameter choices reflect the different roles of the figures and
do not correspond to different physical models.

\subsubsection{Device level simulation}

In contrast to the collision-based reservoir framework, this section addresses a direct device level approach where the $H \varphi H$ sequence is modeled as a driven evolution of a phenomenological two-level transmon qubit approximation under Gaussian-modulated control pulses~\cite{alexander_qiskit_2020, li_pulse-level_2022}. The simulation employs realistic circuit parameters and decoherence constants in order to generate the pre-measurement density matrix. The resulting state is then used in the Fisher information expression, where
derivatives with respect to the encoded parameter are obtained numerically
through finite-difference evaluation. This framework enables a structural comparison with the collision model by highlighting how sensitivity emerges from hardware-level dynamics rather than from sequential reservoir interactions.

We begin by considering the implementation of the Hadamard gate in a form consistent with the native compilation adopted in current IBMQ devices~\cite{ibm_q_experience_2025}. In this framework the operation is decomposed as $H = R_z[\tfrac{\pi}{2}]\,R_x[\tfrac{\pi}{2}]\,R_z[\tfrac{\pi}{2}]$. Since both $R_z$ rotations and the intermediate parameter shift gate $\varphi$ can be realized as virtual frame updates~\cite{ibm_q_experience_2025}, they do not introduce additional noise. Consequently, the only physically driven operation in this sequence is the $R_x[\pi/2]$ rotation, which is implemented as a single qubit microwave pulse. In order to reflect realistic control conditions, we model this drive by a Gaussian-modulated envelope.

Within this description the qubit is treated as an effective two-level transmon, driven by a classical oscillatory electric field. The dynamics are described by the Hamiltonian
\begin{equation}
H(t) = \frac{\hbar \omega_0}{2}\,\sigma_z + \Omega(t)\cos(\omega_D t)\,\sigma_x,
\end{equation}
where $\omega_0$ is the transition frequency of the qubit, $\omega_D$ denotes the drive frequency, and $\Omega(t)$ is the time-dependent Rabi frequency. This choice captures the essential features of experimental single qubit control without invoking further corrections such as higher-level leakage~\cite{hyyppa_reducing_2024, turkpence_anharmonicity_2026}, which are beyond the present scope.
We emphasize that the device-level model used here is an effective two-level
transmon approximation. The Lindblad simulation includes relaxation and
dephasing within the computational subspace, but it does not include higher
transmon levels. Consequently, leakage to non-computational states, finite
anharmonicity effects, pulse-induced population of the second excited state,
calibration errors, pulse distortions, frequency drift, and multi-qubit
crosstalk are not included in the present description. This approximation is
chosen deliberately in order to isolate the phase-sensitivity imprint of the
single-qubit $H\,\varphi\,H$ primitive and to compare it with the reduced
repeated-interaction model without introducing additional device-specific
parameters. A fully calibrated multi-level transmon treatment, including
leakage and crosstalk, is left for future hardware-specific extensions.

 The modulation is chosen as
\begin{equation}
\Omega(t) = \Omega_0 \exp\!\left(-\frac{(t-t_c)^2}{\sigma_p^2}\right),
\end{equation}
with $\Omega_0$ the peak amplitude, $t_c$ the temporal center of the pulse, and $\sigma_p$ the width of the Gaussian envelope. The amplitude parameter $\Omega_0$ specifies the maximum Rabi frequency reached during the Gaussian pulse and is given by 
$\Omega_0 = \frac{\alpha}{\sqrt{\pi}\,\sigma_p}$, 
with the derivation provided in Appendix~\ref{App:D}. 
Here, $\alpha = \int_{-\infty}^{\infty}\Omega(t)\,dt$ denotes the dimensionless pulse area, which directly corresponds to the rotation angle of the Bloch vector, thereby quantifying the net qubit rotation induced by the drive.

In the device–level model, virtual $z$–axis phase updates $R_z[\alpha]$ are treated as ideal frame rotations, whereas the $x$–axis quarter–turns $R_x[\tfrac{\pi}{2}]$ are realized by a Gaussian–modulated drive and therefore evolve under the open–system dynamics in Eq.~(\ref{Eq:Lindblad}). 
We define the noiseless phase–update channel as $\mathcal{U}_z[\beta](\rho) = R_z[\beta]\,\rho\, R_z^\dagger[\beta]$ where $\beta$ denotes the applied phase shift, and the noisy $x$–rotation channel as
\begin{equation}
\mathcal{E}_x\!\left[\tfrac{\pi}{2}\right](\rho) =
\mathcal{T}\exp\!\Big(\int_0^{T_x}\!\mathcal{L}(t)\,dt\Big)\rho,
\end{equation}
with $\mathcal{L}(t)$ the time–dependent Liouvillian generated by the control Hamiltonian $H(t)$ and the jump operators 
$L_1=\sigma_-$ and $L_\phi=\sigma_z$ at rates $\Gamma_1=1/T_1$ and $\gamma_\phi=1/T_2-1/(2T_1)$. 

\begin{figure}[t]
\centering
\includegraphics[width=4.0in]{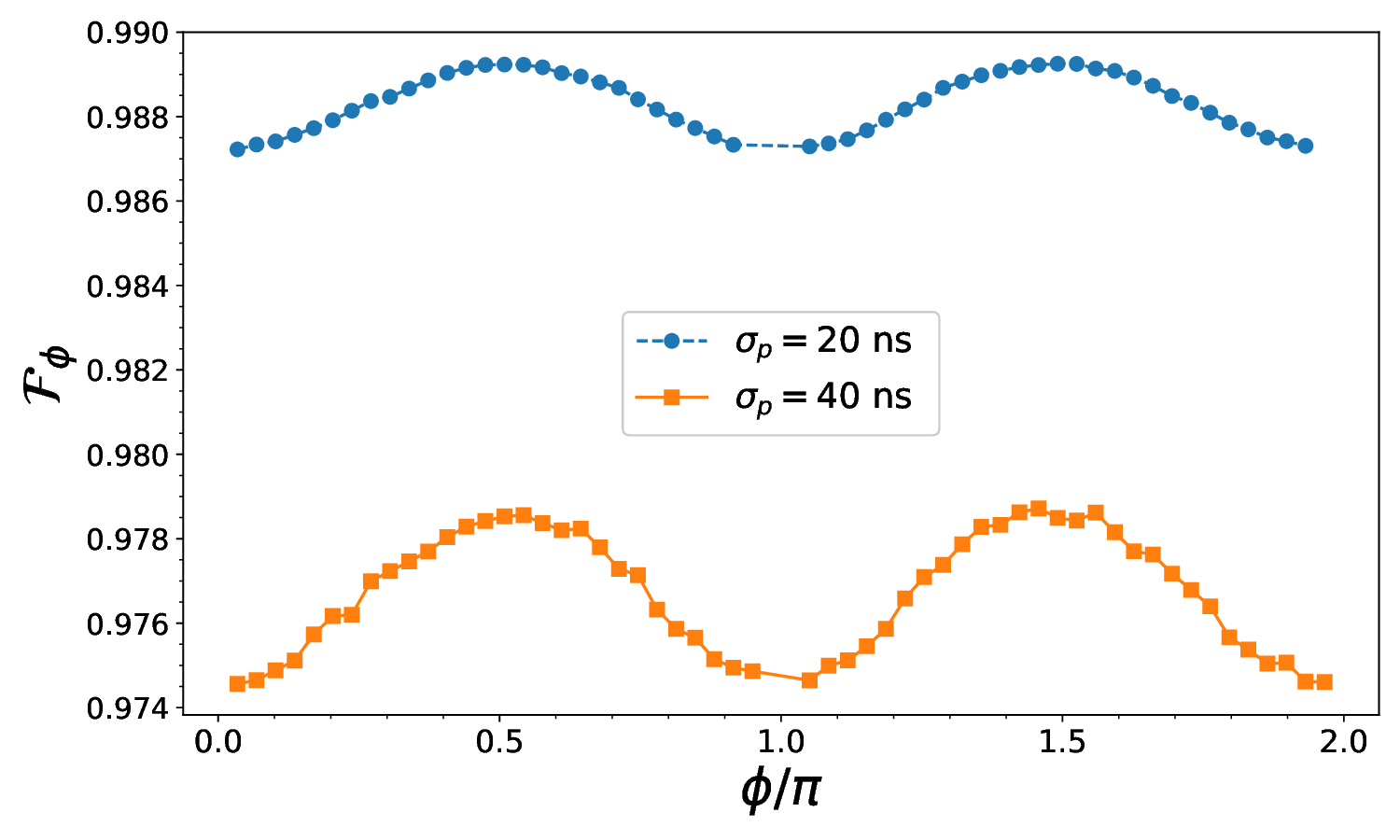} 
\caption{Quantum Fisher information $\mathcal{F}_{\phi}$ obtained from 
two separate device--level simulations of the noisy 
$H\,\varphi\,H$ sequence. 
Each curve corresponds to a distinct implementation of the 
$R_{x}[\pi/2]$ rotations: one sequence uses Gaussian--modulated 
$x$--pulses with width $\sigma_{p}=20~\mathrm{ns}$, and the other 
uses pulses with width $\sigma_{p}=40~\mathrm{ns}$. 
In both cases the pulses satisfy the area condition 
$\alpha=\pi/2$, and the transmon transition frequency is 
$\omega_{0}=2\pi\times 4.5~\mathrm{GHz}$. 
Dissipative effects are modeled with relaxation and dephasing 
constants $T_{1}=150~\mu\mathrm{s}$ and $T_{2}=100~\mu\mathrm{s}$, 
matching the parameters used in the reservoir--based analysis.
}
\label{fig:Fig4}
\end{figure}

The complete channel implementing the noisy $H\,\varphi\,H$ sequence, where
$\varphi=\operatorname{diag}(1,e^{i\phi})$ and $\phi$ denotes
the encoded phase angle, is then
\begin{align}
\Phi_{\mathrm{dev}}(\phi)
= \mathcal{U}_z\!\big[\tfrac{\pi}{2}\big]\circ
  \mathcal{E}_x\!\big[\tfrac{\pi}{2}\big]\circ
  \mathcal{U}_z[\phi+\pi]\circ
  \mathcal{E}_x\!\big[\tfrac{\pi}{2}\big]\circ
  \mathcal{U}_z\!\big[\tfrac{\pi}{2}\big],
\end{align}
so that the output state is
$\rho_{\mathrm{out}}=\Phi_{\mathrm{dev}}(\phi)[\rho_0]$.

We now specify the virtual-$Z$ convention used in this compressed native
sequence. Throughout the device-level simulation, we use
$R_z(\theta)=\operatorname{diag}(e^{-i\theta/2},e^{i\theta/2})$.
With this convention, the phase gate used in the algorithmic sequence is
$\varphi=\operatorname{diag}(1,e^{i\phi})$, which is equivalent to
$R_z(\phi)$ up to a global phase. The Hadamard gate is
implemented, again up to a global phase, as
\begin{equation}
H \equiv R_z(\pi/2)R_x(\pi/2)R_z(\pi/2).
\end{equation}
Consequently, the target sequence $H\,\varphi\,H$ can be written as
\begin{align}
H\,\varphi\,H
&\equiv
R_z(\pi/2)R_x(\pi/2)
R_z(\pi/2)R_z(\phi)R_z(\pi/2)
R_x(\pi/2)R_z(\pi/2) \nonumber\\
&=
R_z(\pi/2)R_x(\pi/2)
R_z(\phi+\pi)
R_x(\pi/2)R_z(\pi/2),
\end{align}
where only physically irrelevant global phases have been omitted. Thus, the
central virtual-$Z$ update $\mathcal{U}_z[\phi+\pi]$ is not an independent
redefinition of the phase gate, but the result of merging the two neighboring
$R_z(\pi/2)$ frame updates with the intended phase rotation. This convention
ensures that the relative phase encoded by the compressed device-level
sequence is the desired value $\phi$.

For consistency with the reservoir-based analysis, we use the same dissipation
constants as in Fig.~\ref{fig:Fig2}, namely $T_1=150~\mu\mathrm{s}$ and
$T_2=100~\mu\mathrm{s}$.

The quantum Fisher information obtained from the device--level simulation is
shown in Fig.~\ref{fig:Fig4}.  
The two curves correspond to $H\,\varphi\,H$ sequences implemented with 
Gaussian $R_{x}[\pi/2]$ rotations of duration $\sigma_{p}=20~\mathrm{ns}$ and 
$\sigma_{p}=40~\mathrm{ns}$, respectively.  
The shorter pulse produces a higher peak Rabi rate and completes the driven 
rotation over a reduced time window, which limits the exposure of the qubit to 
the dissipative channels.  
As a result, the corresponding QFI curve exhibits a larger overall amplitude
and a smoother dependence on~$\phi$, whereas the longer pulse, being subject
to decoherence for nearly twice as long, yields a reduced QFI and a more
pronounced modulation. 
This behavior directly reflects how the total duration of the physical 
$R_{x}[\pi/2]$ operations influences the preserved coherence at the end of 
the $H\,\varphi\,H$ sequence.

The two pulse widths also map naturally onto the effective noise exposure 
intervals used in the analytical reservoir analysis.  
A sequence built from $\sigma_{p}=20~\mathrm{ns}$ pulses occupies an overall 
driven window of approximately $200~\mathrm{ns}$, while 
$\sigma_{p}=40~\mathrm{ns}$ pulses extend this window to roughly 
$400~\mathrm{ns}$.  
This correspondence enables a qualitative comparison with the steady state 
prediction shown in Fig.~\ref{fig:Fig3}, where the same pair of effective 
durations was used as input to the dissipative collision model.

Both approaches exhibit the same hierarchy: the channel with the shorter
effective duration achieves the higher QFI.  
However, the relative separation between the two curves is smaller in 
Fig.~\ref{fig:Fig3}.  
This difference arises from the structural distinction between the models.  
In the device--level simulation, the qubit undergoing the 
$H\,\varphi\,H$ sequence experiences continuous $T_{1}$ and $T_{2}$ processes 
throughout the physically driven rotations.  In contrast, in the collision model the probe qubit itself is not subjected to an explicit external noise channel during the interaction steps; instead, dissipative effects enter indirectly through the prepared ancillas that constitute the reservoir. As a consequence, the parametric imprint is attenuated more weakly in the analytical treatment, whereas the hardware--level simulation incorporates the full temporal exposure of the driven qubit to relaxation and dephasing. 

Despite these differing microscopic mechanisms, the qualitative agreement 
between Fig.~\ref{fig:Fig3} and Fig.~\ref{fig:Fig4} is notable: both predict 
enhanced sensitivity near $\phi=\pi/2$ and $3\pi/2$, reduced sensitivity 
near $\phi=\pi$, and a consistent ordering of QFI amplitudes determined by 
the effective noise duration.  
This alignment indicates that the collision model captures the same
qualitative phase sensitivity structure as the noisy hardware primitive,
even though it abstracts away the detailed driven dynamics of the device.

It is worth noting that the device-level analysis presented here relies on the pre-measurement density matrix obtained from numerical integration of the open-system master equation. We do not execute the protocol on actual cloud-based quantum processors or on classical noisy emulators such as fake backends, since these platforms primarily provide sampled output distributions after measurement. Although such executions are valuable for assessing finite-shot statistics and readout-level performance, extracting the full density matrix would require an additional quantum state tomography layer. This reconstruction step would introduce finite-shot sampling noise, readout errors, and tomography-related artifacts, which would obscure the intrinsic pre-measurement phase-sensitivity profile considered in this work. For this reason, the present comparison uses a controlled density-matrix-based pulse simulation to isolate the analytical structure of $\mathcal{F}\sb{\phi}$, while experimental or backend-level validation based on full tomographic reconstruction is left for future work.

In this work, the term ``tomography-free'' is used in a model-based sense. It
does not refer to an experimental procedure for reconstructing an unknown
quantum state without tomography. Rather, it means that, once the effective
noise parameters and the gate protocol are specified, the phase sensitivity can
be evaluated directly from the analytically or numerically obtained
pre-measurement density matrix, without carrying out an additional experimental
state-reconstruction step. The framework should therefore be understood as a
theoretical diagnostic and reduced-description tool, while full experimental
validation would require a separate tomographic reconstruction protocol.

\subsubsection{Quantitative comparison of phase-sensitivity profiles}

\begin{figure}[t]
\centering
\includegraphics[width=\linewidth]{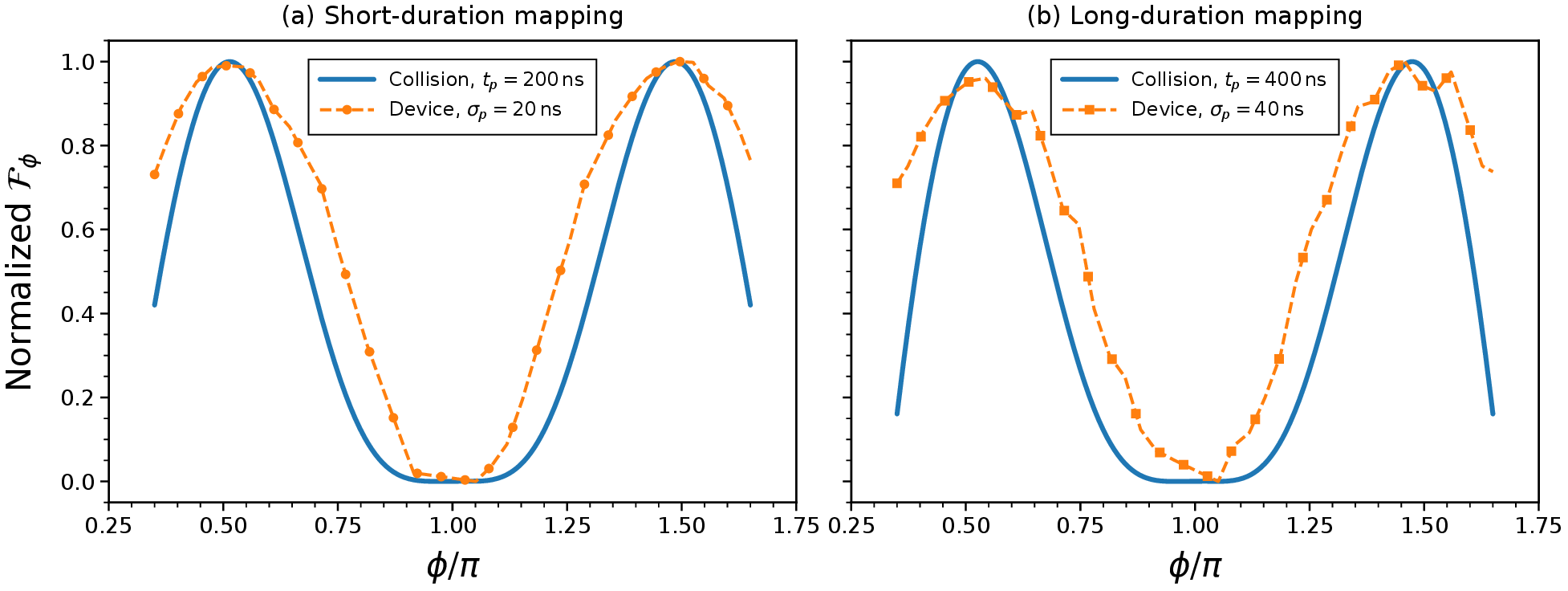}
\caption{\label{fig:Fig5}
Normalized quantum Fisher information profiles used for the quantitative
comparison between the repeated-interaction model and the device-level pulse
simulation. The normalization is performed independently for each curve
according to Eq.~\eqref{Eq:QFI_norm}.
(a) Comparison between the collision-model result for $t_p=200~\mathrm{ns}$
and the device-level result for $\sigma_p=20~\mathrm{ns}$.
(b) Comparison between the collision-model result for $t_p=400~\mathrm{ns}$
and the device-level result for $\sigma_p=40~\mathrm{ns}$.
The comparison is shown over the central phase window
$0.35\leq \phi/\pi \leq 1.65$. Solid curves denote the repeated-interaction
model, while dashed curves with markers denote the device-level simulations.}
\end{figure}

To quantify the correspondence between the repeated-interaction model and the
device-level pulse simulation, we compare the normalized phase-sensitivity
profiles shown in Fig.~\ref{fig:Fig5}. The comparison is performed over the
central phase window $0.35 \leq \phi/\pi \leq 1.65$, which contains the two
dominant maxima and the central minimum of the QFI profile. The device-level
data are interpolated onto the same phase grid used for the collision-model
calculation. The quantities used below are standard descriptive measures for comparing one-dimensional profiles and are used here only to quantify the similarity of the phase-sensitivity landscapes \cite{press_numerical_2007}.

For each curve, the normalized quantum Fisher information is defined as
\begin{equation}
\tilde{\mathcal{F}}_{\phi} =
\frac{\mathcal{F}_{\phi}-\mathcal{F}_{\min}}
{\mathcal{F}_{\max}-\mathcal{F}_{\min}} .
\label{Eq:QFI_norm}
\end{equation}

Here, $\mathcal{F}_{\min}$ and $\mathcal{F}_{\max}$ denote the minimum and maximum values of $\mathcal{F}_{\phi}$ within the comparison window. This normalization removes the difference in absolute QFI scale and allows the
comparison to focus on the phase dependence of the sensitivity profile. We use
the locations of the left maximum, central minimum, and right maximum, extracted
from the phase windows $0.35\leq\phi/\pi\leq0.75$,
$0.75\leq\phi/\pi\leq1.25$, and $1.25\leq\phi/\pi\leq1.65$, respectively.

The corresponding phase-position mismatch is defined as
\begin{equation}
\Delta_{\alpha}
=
\left|
\frac{\phi_{\alpha}^{\mathrm{coll}}}{\pi}
-
\frac{\phi_{\alpha}^{\mathrm{dev}}}{\pi}
\right|,
\qquad
\alpha=L,V,R .
\label{Eq:Delta_peak}
\end{equation}
Here, $L$, $V$, and $R$ denote the left maximum, central valley, and right maximum, respectively.

We also compute the peak-to-valley contrast as
\begin{equation}
C\sb{\mathrm{PV}}
=
\frac{
0.5\left(\mathcal{F}\sb{L}\sp{\max}+\mathcal{F}\sb{R}\sp{\max}\right)
-
\mathcal{F}\sb{V}\sp{\min}
}
{\mathcal{F}\sb{\mathrm{mean}}}.
\label{Eq:CPV}
\end{equation}
Here, $\mathcal{F}\sb{L}\sp{\max}$ and $\mathcal{F}\sb{R}\sp{\max}$ are the QFI values at the left and right maxima, $\mathcal{F}\sb{V}\sp{\min}$ is the QFI value at the central minimum, and $\mathcal{F}\sb{\mathrm{mean}}$ denotes the mean QFI over the comparison window.

To compare the normalized curve shapes, we define
$f\sb{j}\sp{\mathrm{coll}}=\tilde{\mathcal{F}}\sb{\phi,j}\sp{\mathrm{coll}}$
and
$f\sb{j}\sp{\mathrm{dev}}=\tilde{\mathcal{F}}\sb{\phi,j}\sp{\mathrm{dev}}$
on a common grid of $N$ phase points. The normalized curve distance is then
\begin{equation}
D\sb{2}
=
\left[
\frac{1}{N}
\sum\sb{j=1}\sp{N}
\left(
f\sb{j}\sp{\mathrm{coll}}
-
f\sb{j}\sp{\mathrm{dev}}
\right)\sp{2}
\right]\sp{1/2}.
\label{Eq:D2}
\end{equation}

Finally, the Pearson correlation coefficient is computed as
\begin{align}
R
&=
\frac{
\sum_{j=1}^{N}
\left(f_j^{\mathrm{coll}}-\bar{f}^{\mathrm{coll}}\right)
\left(f_j^{\mathrm{dev}}-\bar{f}^{\mathrm{dev}}\right)
}{
\sqrt{
\sum_{j=1}^{N}
\left(f_j^{\mathrm{coll}}-\bar{f}^{\mathrm{coll}}\right)^2
}
\sqrt{
\sum_{j=1}^{N}
\left(f_j^{\mathrm{dev}}-\bar{f}^{\mathrm{dev}}\right)^2
}
},
\label{Eq:Rcorr}
\end{align}
where $\bar{f}^{\mathrm{coll}}$ and $\bar{f}^{\mathrm{dev}}$ denote the mean
values of the normalized collision-model and device-level profiles over the
comparison window.

The resulting quantitative measures are summarized in
Table~\ref{tab:quantitative_comparison}.For the short-duration pair, the mismatch of the two maxima is below $0.008\pi$, while the central-valley mismatch is $0.0156\pi$. The corresponding normalized curve distance is $D\sb{2}=0.1717$, and the correlation coefficient is $R=0.9564$. For the long-duration pair, the two
peak mismatches are both $0.0156\pi$, while the central-valley mismatch is
$0.0026\pi$. In this case, the normalized curve distance is $D\sb{2}=0.1953$,
and the correlation coefficient is $R=0.9353$. The high values of $R$ in both
cases indicate a strong structural agreement between the normalized
phase-sensitivity profiles.

\begin{center}
\refstepcounter{table}
\label{tab:quantitative_comparison}

\textbf{Table \thetable.}
Quantitative comparison between the repeated-interaction model and the
device-level pulse simulation over the phase window
$0.35\leq\phi/\pi\leq1.65$. The quantities $\Delta\sb{L}$,
$\Delta\sb{V}$, and $\Delta\sb{R}$ denote the absolute mismatch in the
locations of the left maximum, central minimum, and right maximum,
respectively, in units of $\phi/\pi$. The quantities
$C\sb{\mathrm{PV}}\sp{\mathrm{coll}}$ and
$C\sb{\mathrm{PV}}\sp{\mathrm{dev}}$ denote the peak-to-valley contrast of
the collision-model and device-level curves.

\vspace{2mm}

\small

\begin{tabular}{
l
c
c
c
c
c
c
c
}
\hline
Case &
$\Delta\sb{L}$ &
$\Delta\sb{V}$ &
$\Delta\sb{R}$ &
$C\sb{\mathrm{PV}}\sp{\mathrm{coll}}$ &
$C\sb{\mathrm{PV}}\sp{\mathrm{dev}}$ &
$D\sb{2}$ &
$R$
\tabularnewline
\hline
Short-duration pair &
0.0052 &
0.0156 &
0.0078 &
$1.2973\times 10\sp{-2}$ &
$1.9731\times 10\sp{-3}$ &
0.1717 &
0.9564
\tabularnewline
Long-duration pair &
0.0156 &
0.0026 &
0.0156 &
$1.1753\times 10\sp{-2}$ &
$4.0859\times 10\sp{-3}$ &
0.1953 &
0.9353
\tabularnewline
\hline
\end{tabular}

\end{center}

The difference in peak-to-valley contrast provides a quantitative measure of
the remaining model-dependent discrepancy. In particular, the device-level
simulation gives $C\sb{\mathrm{PV}}\sp{\mathrm{dev}}\sim 10\sp{-3}$, whereas
the repeated-interaction model gives
$C\sb{\mathrm{PV}}\sp{\mathrm{coll}}\sim 10\sp{-2}$. This difference is
consistent with the fact that, in the device-level simulation, the qubit is
continuously exposed to relaxation and dephasing during the finite pulse
sequence, whereas in the repeated-interaction description the noise enters
indirectly through the prepared ancilla states and the subsequent partial-trace
dynamics.

These results show that the two descriptions identify the same phase-sensitive
regions, even though they are not microscopically equivalent. The
repeated-interaction model and the device-level simulation therefore agree at
the level of the normalized phase-sensitivity landscape, while differences in
the absolute QFI scale and contrast reflect the distinct physical mechanisms
included in the two descriptions.

\section{Implications and Outlook}\label{Sect-imp}

The results presented here illustrate a model-based, tomography-free route for
characterizing the phase-sensitivity profile $\mathcal{F}_{\phi}$ of a
device's pre-measurement quantum state. We emphasize that the following
discussion should be understood as an outlook on possible applications of such
phase-sensitivity information, rather than as a direct demonstration of quantum
error correction (QEC), stabilizer optimization, or compiler-level improvement
within the present work. The main result established here is the identification
of phase-dependent regions of enhanced and reduced QFI in analytically
tractable repeated-interaction dynamics and in a corresponding device-level
open-system simulation.

One possible future direction concerns QEC settings in which syndrome
extraction can be interpreted as a metrological task operating under quantum
noise constraints~\cite{barends_superconducting_2014, acharya_quantum_2025,
roffe_quantum_2019, akahoshi_partially_2024}. In such settings, knowledge of
how $\mathcal{F}_{\phi}$ varies with the encoded phase could be used as a
diagnostic input for future studies of hardware-aware syndrome-measurement
strategies, particularly in biased-noise architectures where phase-flip errors
play a central role~\cite{huang_between_2021, bonilla_ataides_xzzx_2021}.
For example, phase regions exhibiting enhanced sensitivity could be examined
as candidates where stabilizer choices or measurement bases require additional
robustness, whereas regions of reduced sensitivity could be investigated as
potential operating points for phase-noise resilience~\cite{len_quantum_2022,
zhou_optimal_2023}. These possibilities remain prospective and require
dedicated QEC-level simulations before quantitative conclusions can be drawn.

A second possible direction concerns compiler and pulse-level
optimization~\cite{alexander_qiskit_2020, li_pulse-level_2022}. The phase-sensitivity profile obtained here could serve
as a diagnostic map for future investigations of whether gate decompositions
or pulse schedules can be arranged to avoid highly phase-sensitive regions.
Similarly, pulse-engineering techniques, including DRAG-type shaping, could be
tested in future work to determine whether they suppress phase-fragile dynamics
identified by peaks in $\mathcal{F}_{\phi}$~\cite{chow_optimized_2010,
hyyppa_reducing_2024, motzoi_improving_2013}. For multi-qubit architectures,
spatial variations of such phase-sensitivity profiles across a device could
also be explored as a possible input for hardware-aware placement of
phase-critical algorithmic layers. These compiler-level and pulse-level
implications are not claimed as demonstrated outcomes of the present study,
but rather as natural directions enabled by the diagnostic information provided
by the QFI landscape.

A further implication is that the collision-model description provides an
analytically transparent and computationally lightweight complement to
full pulse-level simulations. Its closed-form steady-state expressions enable
rapid parameter sweeps and controlled variation of noise models, facilitating
systematic exploration of the roles of $T_{1}$, $T_{2}$, interaction strength
$g$, and reservoir structure. The comparison with device-level simulations
suggests that this simplified framework can be useful as a theoretical
screening tool for exploring noise-aware protocol design and
reservoir-engineered primitives for near-term quantum hardware, provided that
its predictions are subsequently validated with more detailed device-specific
models.

Although our analysis focused on a single-qubit probe, the approach naturally
extends to composite probes and multi-qubit reservoirs~\cite{cattaneo_collision_2021}.
Entangled probes could be used in future work to test whether the qualitative
agreement between collision-model predictions and device-level simulations
persists beyond the single-qubit regime. Likewise, correlated or structured
reservoirs could reveal how environmental correlations modify steady-state
phase imprinting. These directions suggest a broader program in which
collision models and pulse-level simulations jointly support the development
of noise-aware and hardware-compatible quantum technologies.

\section{CONCLUSIONS}\label{Sect-conc}

In this work, we have examined the phase sensitivity of noisy quantum devices
from two complementary perspectives.
Closed-form analytical expressions were derived within a collision model
framework and compared with pulse-resolved open-system simulations of a noisy
$H,\varphi,H$ sequence.
Across these two distinct dynamical descriptions, the quantum Fisher
information $\mathcal{F}_{\phi}$ was found to exhibit a consistent
phase-dependent structure.
The analytical steady-state solution shows that the probe qubit
can carry measurable metrological information about an algorithmically prepared
reservoir, even in the presence of dissipation and dephasing. The physical
origin of this metrological response is the persistence of steady coherence in
the asymptotic probe state. In particular, the finite off-diagonal elements of
$\rho\sb{S}\sp{\mathrm{ss}}(\phi)$ provide the channel through which the phase
imprint of the prepared ancillas is retained by the probe.

The agreement with the pulse-level simulation should therefore
be understood as supporting evidence that the same algorithmic phase imprint
can be reflected in different dynamical descriptions. In the
repeated-interaction model, this imprint is mediated by finite-time prepared
ancillas and partial tracing. In the device-level simulation, it emerges from
continuous noisy evolution during the driven $H,\varphi,H$ pulse sequence.
The quantitative comparison added in Section IV shows that, despite these
different mechanisms, the two descriptions identify the same phase-sensitive
regions of the QFI landscape. This provides a concrete link between an
analytically tractable repeated-interaction model and a pulse-level open-system
description of a noisy single-qubit phase-encoding circuit. In this sense, the
main contribution of the work is not a new collision model in isolation, but a
reduced diagnostic framework that maps algorithmically prepared noisy ancillas
to the phase-sensitivity structure observed in a device-level implementation.

Beyond conceptual insight, the framework developed here provides a practical
route for characterizing phase sensitivity in near-term hardware without
relying on full state tomography.
Because $\mathcal{F}_{\phi}$ is obtained directly from the pre-measurement
state, it captures the intrinsic phase response of a device in a manner that
is closely tied to realistic noise processes.
The resulting $\mathcal{F}_{\phi}$ profiles identify operating regions of
enhanced or reduced phase sensitivity, thereby offering hardware-informed
guidance for error-mitigation strategies, stabilizer choices, and
compiler-level gate synthesis.

Finally, the analytical tractability of the collision model makes it a useful
theoretical platform for exploring more complex scenarios, including
structured reservoirs, non-Markovian effects, and entangled probe systems.
Extending this approach to multi-qubit settings may help clarify how
environmental correlations influence parameter sensitivity in larger
processors and support the development of reservoir-engineered and
noise-aware protocols for future quantum technologies.

%
%

\ack{This work was supported by the Scientific and Technological Research Council of T\"{u}rkiye (T\"{U}B{\.I}TAK, Grant No.\ 125F473). 
We also acknowledge the facilities and technical support provided by the Informatics Institute of \.{I}stanbul Technical University 
and the Qready Quantum Technologies and Consulting Joint Stock Company.}


\data{Data supporting this study are available from the corresponding author upon reasonable request.}

\appendix
\section{Derivation of the micromaser-type master equation}\label{App:A}

This appendix contains the complete algebraic derivation of the Lindblad master equation quoted in Sec.~\ref{Sect-model}. We work in units where $\hbar=1$. 
Before each collision the joint state factorizes as $\rho(t)=\rho_S(t)\otimes\rho'$. 
The propagator for a single collision of duration $\tau$ is expanded to second order,
\begin{align}
U(\tau) &\simeq \mathbbm{1} - V_1(\tau) - V_2(\tau), \\
V_1(\tau) &= i\tau H_{\mathrm{int}}, \qquad
V_2(\tau) = \frac{\tau^2}{2} H_{\mathrm{int}}^2.
\end{align}

The validity of the above second-order micromaser expansion is controlled by
the dimensionless interaction parameter accumulated during a single collision,
namely $g\tau$. The approximation requires $g\tau\ll 1$, so that terms of
third and higher order in the short-time expansion of the interaction-picture
propagator remain negligible. This condition is satisfied by the numerical
parameter sets used in the manuscript. For the mutual-information dynamics,
we use $g=0.10$ and $\tau=0.12$, which gives
$g\tau=0.012$ and the nominal cubic scale
$(g\tau)\sp{3}\simeq1.7\times10\sp{-6}$. For the QFI profiles, we use
$g=0.15$ and $\tau=0.80$, which gives $g\tau=0.12$ and
$(g\tau)\sp{3}\simeq1.7\times10\sp{-3}$. Therefore, the largest value of
$g\tau$ considered here remains well below unity, and the calculations are
performed within the perturbative regime assumed in deriving the micromaser
master equation.

Ancilla–probe interactions are assumed to occur randomly according to a
Poisson process with rate $r$. Over a short time interval $\delta t$,
the reduced state of the probe evolves as
\begin{align}
\rho_S(t+\delta t)
= (1-r\delta t)\,\rho_S(t) + r\delta t \,
\mathrm{Tr}_{R}\!\left[U(\tau)\,(\rho_S\otimes\rho')\,U^{\dagger}(\tau)\right].
\end{align}
Subtracting $\rho_S(t)$ from both sides, dividing by $\delta t$, and
taking the limit $\delta t\to 0$, one obtains
\begin{align}
\dot{\rho}_S(t)
= r\left(
\mathrm{Tr}_{R}\!\left[
U(\tau)\,(\rho_S(t)\otimes\rho')\,U^{\dagger}(\tau)
\right]
-\rho_S(t)
\right).
\label{Eq:CollisionME}
\end{align}
In the numerical and analytical evaluations presented in this work, the
collision rate is fixed to $r=1$, which corresponds to a rescaling of the
time variable and does not affect the qualitative structure of the
steady state dynamics.

Inserting the second-order expansion and discarding $\mathcal{O}(\tau^3)$ terms, the first-order part yields the effective Hamiltonian
\begin{align}
H_{\mathrm{eff}}
= r\,g\,\tau\!\left(
\langle\sigma^{-}\rangle_{\rho'}\,\sigma^{+}_{S}
+
\langle\sigma^{+}\rangle_{\rho'}\,\sigma^{-}_{S}
\right),
\end{align}
while the second-order part produces dissipation. 
After tracing over the ancilla one obtains

\begin{align}
\mathrm{Tr}_{R}\Bigl[V_1(\rho_S\otimes\rho')V_1^{\dagger}
- \tfrac{1}{2}\{H_{\mathrm{int}}^2,\rho_S\otimes\rho'\}\Bigr] 
&= \frac{\tau^{2}g^{2}}{2}
\Bigl[
2\langle\sigma^{+}\sigma^{-}\rangle_{\rho'}\,\sigma^{+}_{S}\rho_S\sigma^{-}_{S}
- \langle\sigma^{+}\sigma^{-}\rangle_{\rho'}\,\{\sigma^{+}_{S}\sigma^{-}_{S},\rho_S\}\nonumber\\
&+ 2\langle\sigma^{-}\sigma^{+}\rangle_{\rho'}\,\sigma^{-}_{S}\rho_S\sigma^{+}_{S}
- \langle\sigma^{-}\sigma^{+}\rangle_{\rho'}\,\{\sigma^{-}_{S}\sigma^{+}_{S},\rho_S\}
\Bigr].
\end{align}
Using the standard Lindblad form $\mathcal{L}[o]\rho = 2o\rho o^{\dagger} - \{o^{\dagger}o,\rho\}$, the full master equation becomes
\begin{align}
\dot{\rho}_S
&= -i[H_{\mathrm{eff}},\rho_S]
 + \Gamma_{+}\,\mathcal{L}[\sigma^{+}_{S}]\rho_S
 + \Gamma_{-}\,\mathcal{L}[\sigma^{-}_{S}]\rho_S,
\end{align}
with the rates given in the main text. 
This derivation reproduces exactly Eqs.~(12)--(15).

\section{Explicit steady state density matrix}\label{App:B}

The noisy ancilla emerging from the $H\,\varphi\,H$ gate sequence is described by the state
\begin{equation}\label{Eq:Rho_noise_corrected}
\rho'(\phi) =
\begin{pmatrix}
\displaystyle \frac{1+\cos\phi}{2}\,e^{-\gamma_1 t_p} 
  & \displaystyle \frac{i\sin\phi}{2}\,e^{-\gamma_2 t_p} \\[6pt]
\displaystyle -\frac{i\sin\phi}{2}\,e^{-\gamma_2 t_p} 
  & \displaystyle 1 - \frac{1+\cos\phi}{2}\,e^{-\gamma_1 t_p}
\end{pmatrix},
\end{equation}
where $\gamma_1 = 1/T_1$ and $\gamma_2 = 1/T_2$ denote the relaxation and dephasing rates, and $t_p$ denotes the finite noise--preparation interval during which each reservoir ancilla is exposed to dissipation prior to its interaction with the probe. Importantly, $t_p$ is a fixed, finite preparation time and does not represent a long time or asymptotic limit.

The expectation values entering the micromaser-type generator follow directly from Eq.~(\ref{Eq:Rho_noise_corrected}):
\begin{align}
\langle \sigma^+\sigma^- \rangle_{\rho'} &= \frac{1+\cos\phi}{2}\,e^{-\gamma_1 t_p} \equiv p_0, \\[2pt]
\langle \sigma^-\sigma^+ \rangle_{\rho'} &= 1 - p_0, \\[2pt]
\langle \sigma^- \rangle_{\rho'} &= \frac{i\sin\phi}{2}\,e^{-\gamma_2 t_p}.
\end{align}
Introducing the shorthand $\zeta \equiv r\tau g$ and $\Gamma_{12}=\gamma_1+\gamma_2$, and substituting these averages into the general steady state form of Eq.~(\ref{Eq:Steady_rho}), one obtains the compact steady state probe density matrix
\begin{align}\label{Eq:Steady_rho_corrected}
\rho_S^{\mathrm{ss}}(\phi) =
\begin{pmatrix}
p_0 & \rho_{S,01}^{\mathrm{ss}} \\[8pt]
(\rho_{S,01}^{\mathrm{ss}})^{*} & 1-p_0
\end{pmatrix}.
\end{align}

The off-diagonal element is determined by the imbalance of the ancilla excitation probabilities and the residual coherence of $\rho'(\phi)$:
\begin{align}\label{Eq:Steady_rho_01_final}
\rho_{S,01}^{\mathrm{ss}}
= \frac{\zeta}{2}\,\sin\phi
\left[
e^{-\gamma_2 t_p} - (1+\cos\phi)e^{-\Gamma_{12} t_p}
\right].
\end{align}

The diagonal entries satisfy $\rho_{S,00}^{\mathrm{ss}} = p_0$ and $\rho_{S,11}^{\mathrm{ss}} = 1-p_0$, ensuring that $\operatorname{Tr}(\rho_S^{\mathrm{ss}})=1$. 
The steady state $\rho_S^{\mathrm{ss}}(\phi)$ arises as the asymptotic fixed point of the probe dynamics under repeated interactions with identically prepared ancillas, while each ancilla state $\rho'(\phi)$ itself is generated at a finite preparation time $t_p$ and is not associated with a long time relaxation limit.

\section{Quantum Fisher information for the steady state}\label{App:C}

In this appendix we compute explicitly the quantum Fisher information (QFI) associated with the steady state probe qubit derived in Appendix~\ref{App:B}. 
For a qubit written in Bloch form,

\begin{equation}
\rho_S^{\mathrm{ss}}(\phi)
= \frac{1}{2}\bigl(\mathbb{I} + \boldsymbol{r}(\phi)\cdot\boldsymbol{\sigma}\bigr),
\qquad
\boldsymbol{r}(\phi) = \bigl(r_x(\phi),\,r_y(\phi),\,r_z(\phi)\bigr),
\end{equation}
the QFI for the parameter $\phi$ is given by the Bloch vector expression

\begin{equation}
\mathcal{F}_\phi 
= \bigl|\partial_\phi \boldsymbol{r}\bigr|^2 
\;+\; 
\frac{\bigl(\boldsymbol{r}\cdot \partial_\phi \boldsymbol{r}\bigr)^2}
     {1-|\boldsymbol{r}|^2}.
\end{equation}

The components of the Bloch vector follow from the standard identities
\[
r_x = \mathrm{Tr}\!\left[\rho_S^{\mathrm{ss}}\sigma_x\right], \qquad
r_y = \mathrm{Tr}\!\left[\rho_S^{\mathrm{ss}}\sigma_y\right], \qquad
r_z = \mathrm{Tr}\!\left[\rho_S^{\mathrm{ss}}\sigma_z\right],
\]
which, using the steady state matrix in Eq.~(\ref{Eq:Steady_rho_corrected}), yield
\begin{align}
r_x(\phi)
&= \zeta\,\sin\phi\Bigl(e^{-\gamma_2 t_p} - (1+\cos\phi)e^{-\Gamma_{12} t_p}\Bigr), \\[4pt]
r_y(\phi) &= 0, \\[4pt]
r_z(\phi)
&= (1+\cos\phi)e^{-\gamma_1 t_p} - 1 .
\end{align}

The derivatives of these components with respect to $\phi$ are
\begin{align}
\partial_\phi r_x(\phi)
&= \zeta\,\cos\phi\Bigl(e^{-\gamma_2 t_p} - (1+\cos\phi)e^{-\Gamma_{12} t_p}\Bigr)\nonumber
\\
&\quad {}+ \zeta\,\sin^2\phi\,e^{-\Gamma_{12} t_p}, \\[6pt]
\partial_\phi r_y(\phi) &= 0, \\[6pt]
\partial_\phi r_z(\phi) &= -\,\sin\phi\,e^{-\gamma_1 t_p}.
\end{align}

The vector norms and scalar products entering the QFI formula evaluate to
\begin{align}
\bigl|\partial_\phi \boldsymbol{r}(\phi)\bigr|^2
&= \bigl[\partial_\phi r_x(\phi)\bigr]^2
   + \bigl[\partial_\phi r_z(\phi)\bigr]^2, \\[6pt]
\boldsymbol{r}(\phi)\cdot\partial_\phi \boldsymbol{r}(\phi)
&= r_x(\phi)\,\partial_\phi r_x(\phi)
 + r_z(\phi)\,\partial_\phi r_z(\phi), \\[6pt]
|\boldsymbol{r}(\phi)|^2
&= r_x^2(\phi) + r_z^2(\phi).
\end{align}

Substituting these expressions into the Bloch vector representation of the QFI gives the closed form
\begin{align}\label{Eq:QFI-closed}
\mathcal{F}_\phi
&= \bigl[\partial_\phi r_x(\phi)\bigr]^2 
 + \bigl[\partial_\phi r_z(\phi)\bigr]^2 \nonumber
\\[4pt]
&\quad {}+ 
\frac{\Bigl(r_x(\phi)\,\partial_\phi r_x(\phi)
      + r_z(\phi)\,\partial_\phi r_z(\phi)\Bigr)^2}
     {1 - r_x^2(\phi) - r_z^2(\phi)}.
\end{align}

\section{Algorithmic simulation}\label{App:D}
\subsection{Device-level pulse simulation and numerical settings}

To define a dimensionless measure of the overall drive, we introduce $\alpha$ as the time integral of the Rabi frequency:
\begin{align}\label{Eq:Alpha}
\alpha &= \int_{-\infty}^{\infty} \Omega(t)\, dt 
       = \frac{\mu}{\hbar} \int_{-\infty}^{\infty} E(t)\, dt .
\end{align}
For a Gaussian drive field with an envelope centered at $t_c$:
\begin{equation}\label{Eq:E_t}
E(t) = A_0 \exp\!\left[-\frac{(t-t_c)^2}{\sigma_p^2}\right],
\end{equation}
the integral evaluates to
\begin{equation}\label{Eq:Alpha_eval}
\alpha = \frac{\mu A_0}{\hbar}\sqrt{\pi}\,\sigma_p .
\end{equation}
Rearranging gives the required peak amplitude of the field:
\begin{equation}\label{Eq:Amplitude}
A_0 = \frac{\alpha\hbar}{\mu\sqrt{\pi}\sigma_p} .
\end{equation}
Since the Rabi frequency is $\Omega(t) = \mu E(t)/\hbar$, the peak Rabi rate $\Omega_0 = \Omega(t_c)$ becomes:
\begin{equation}\label{Eq:Rabi}
\Omega_0 = \frac{\alpha}{\sqrt{\pi}\,\sigma_p}.
\end{equation}

This relation shows that the effective peak Rabi rate $\Omega_0$ used in the device level description is determined solely by the normalized pulse area $\alpha$ and the temporal width $\sigma_p$. We simulate a single transmon operated as an effective two–level system and driven along the $x$ axis by a Gaussian–enveloped pulse within the open–system model of Eq.~(\ref{Eq:Lindblad}).  
Throughout the numerics we set the qubit transition frequency to $\omega_0 = 2\pi\times 4.5~\mathrm{GHz}$, which lies well inside the typical $4$–$6~\mathrm{GHz}$ window for fixed–frequency devices.

The device-level simulations are performed by solving the Lindblad master
equation with the \texttt{mesolve} routine of QuTiP. The evolution is carried
out in the laboratory frame for the Hamiltonian in Eq.~(\ref{Eq:Lindblad}) and
the Gaussian-modulated drive defined above. The collapse operators are chosen
as
\begin{equation}
C\sb{1}=\sqrt{\Gamma\sb{1}},\sigma\sb{-},
\qquad
C\sb{\phi}=\sqrt{\gamma\sb{\phi}},\sigma\sb{z},
\end{equation}
where $\Gamma\sb{1}=1/T\sb{1}$ and
$\gamma\sb{\phi}=1/T\sb{2}-1/(2T\sb{1})$. In the numerical simulations we use
$T\sb{1}=150~\mu\mathrm{s}$ and $T\sb{2}=100~\mu\mathrm{s}$, giving
$\Gamma\sb{1}=6.67\times10\sp{3}~\mathrm{s}\sp{-1}$ and
$\gamma\sb{\phi}=6.67\times10\sp{3}~\mathrm{s}\sp{-1}$.

The carrier frequency is fixed at
$\omega\sb{0}=2\pi\times4.5~\mathrm{GHz}$, corresponding to a carrier period
$T\sb{0}=2\pi/\omega\sb{0}\simeq0.222~\mathrm{ns}$. The time grid is chosen to
resolve both the fast carrier oscillation and the Gaussian envelope. For the
QFI sweeps, the finite simulation window is centered at the pulse maximum and
covers $\pm4\sigma\sb{p}$, i.e. $T=8\sigma\sb{p}$. At the two boundaries the
Gaussian envelope is suppressed by $\exp(-16)$, so that the pulse tails are
negligible on the numerical scale of the calculation. The time grid uses at
least 25 points per carrier period, with a minimum of 4000 points. For
$\omega\sb{0}/2\pi=4.5~\mathrm{GHz}$, this gives a typical time step of
approximately $8.9~\mathrm{ps}$. The master-equation solver is run with absolute and relative tolerances $10\sp{-9}$ and a maximum internal-step count of $5\times10\sp{4}$. As a single-pulse diagnostic, we also verified the driven $R\sb{x}[\pi/2]$ gate with a stricter step cap, $\Delta t=\min(T\sb{0}/20,\sigma\sb{p}/200)$, together with tolerances $\mathrm{atol}=10\sp{-9}$ and $\mathrm{rtol}=10\sp{-11}$. For $\sigma\sb{p}=20~\mathrm{ns}$ this gives $\Delta t=11.11~\mathrm{ps}$ and 18001 time points over a $200~\mathrm{ns}$ window.

In our simulation of the $H\,\varphi\,H$ channel, the implementation of the Hadamard gate requires two $\pi/2$ rotations of the form $R_x[\pi/2]$.  
Each such rotation is generated by enforcing the pulse-area condition $\alpha=\pi/2$.  
The total duration of the driven $R_x[\pi/2]$ gate is chosen to fully capture the Gaussian envelope to numerical accuracy, while keeping spectral broadening under control, as detailed in the numerical settings above. 
We examine two pulse widths, $\sigma_p = 20~\mathrm{ns}$ and $\sigma_p = 40~\mathrm{ns}$.

According to Eq.~(\ref{Eq:Rabi}), the shorter pulse width ($\sigma_p=20~\mathrm{ns}$) requires a higher peak Rabi rate ($\Omega_0 \approx 7.05~\mathrm{MHz}$) in order to maintain $\alpha=\pi/2$.  
This produces a faster physical gate and reduces the time over which the phenomenological decoherence channels act during the preparation of the $H\,\varphi\,H$ state.  
Conversely, the longer pulse ($\sigma_p=40~\mathrm{ns}$) operates with a lower peak Rabi rate ($\Omega_0 \approx 3.53~\mathrm{MHz}$) and therefore experiences these noise channels for a substantially longer time.  
All gate evolutions and trajectories are obtained directly in the laboratory frame.

\subsection{Numerical evaluation of QFI}

In the device-level simulation, the output state of the
noisy $H\,\varphi\,H$ channel is obtained directly from the full
Lindblad time evolution. Rather than working with the density matrix
itself, we evaluate the quantum Fisher information using its equivalent
single-qubit Bloch-vector expression, which is particularly convenient
for numerical implementation.
For a Bloch vector
$\mathbf{r}(\phi)=\mathrm{Tr}\!\left[\rho(\phi)\boldsymbol{\sigma}\right]$,
the QFI takes the form
\begin{align}
\mathcal{F}_{\phi}
=
\left|
\frac{\partial \mathbf{r}}{\partial \phi}
\right|^{2}
+
\frac{
\left(
\mathbf{r}\cdot
\frac{\partial\mathbf{r}}{\partial\phi}
\right)^{2}
}{
1-|\mathbf{r}|^{2}
}.
\end{align}

This representation avoids manipulating the full density matrix and allows
the phase sensitivity to be extracted directly from the numerically
computed Bloch trajectory of the probe qubit. To evaluate the derivative
$\partial_{\phi}\mathbf{r}$, we employ a second-order accurate central
finite-difference approximation
\begin{align}
\frac{\partial\mathbf{r}}{\partial\phi}
\;\approx\;
\frac{\mathbf{r}(\phi+\delta\phi)-\mathbf{r}(\phi-\delta\phi)}
{2\,\delta\phi},
\end{align}
where $\mathbf{r}(\phi\pm\delta\phi)$ are obtained from independent
Lindblad evolutions of the channels $\Phi_{\mathrm{dev}}(\phi\pm\delta\phi)$.

For the reported QFI profiles, the finite-difference step is fixed as
\begin{equation}
\delta\phi=10^{-3}.
\end{equation}
Thus, for each sampled phase value, three independent device-level Lindblad
evolutions are carried out, corresponding to
$\Phi_{\mathrm{dev}}(\phi)$,
$\Phi_{\mathrm{dev}}(\phi+\delta\phi)$, and
$\Phi_{\mathrm{dev}}(\phi-\delta\phi)$. The same QuTiP \texttt{mesolve}
solver settings and time-grid construction described in the previous
subsection are used for all three evolutions, so that the finite-difference
derivative is evaluated under identical numerical conditions.

In practice, a grid of $N_{\phi}=60$ phase points is sufficient to resolve the
oscillatory structure of $\mathcal{F}_{\phi}$, resulting in approximately
$3N_{\phi}$ complete master-equation evolutions.

The phase grid used for the reported curves excludes the exact endpoints by
using
\begin{equation}
\phi\in[\epsilon_{\phi},2\pi-\epsilon_{\phi}],
\qquad
\epsilon_{\phi}=10^{-3}.
\end{equation}
This small offset avoids numerical endpoint artifacts at $0$ and $2\pi$ while
leaving the resolved phase dependence unchanged on the scale of the plotted
profiles.

Finally, to avoid numerical instabilities when the state approaches purity
(i.e.\ $|\mathbf{r}|^{2}\rightarrow1$), the denominator
$1-|\mathbf{r}|^{2}$ in the QFI formula is regularized by imposing a small
floor value. This prevents artificial divergences and ensures that the
computed QFI remains well behaved even when the Bloch vector lies very close
to the surface of the Bloch sphere.

In the numerical implementation, this regularization is applied as
\begin{equation}
1-|\mathbf{r}|^{2}
\;\longrightarrow\;
\max\!\left(1-|\mathbf{r}|^{2},\epsilon_{\mathrm{QFI}}\right),
\qquad
\epsilon_{\mathrm{QFI}}=10^{-12}.
\end{equation}
The floor only becomes active when the numerical state is extremely close to
a pure state and is used solely to suppress round-off-induced divergences in
the second term of the Bloch-vector QFI expression.


\end{document}